\documentclass[conference]{IEEEtran}
\usepackage{graphicx,xcolor,tikz}
\usepackage{comment}
\usepackage{subfig}
\usepackage{float}
\usepackage{multicol}
\usepackage{cite}
\usepackage{makecell}
\usepackage{siunitx}
\usepackage{multirow}
\usepackage{mathtools}
\usepackage{amsmath,amssymb, amsfonts}
\usepackage{tabulary}

\usepackage{cite}

\usepackage{mathtools}

\def\nb0{{\mathbf{0}}}
\def\nb1{{\mathbf{1}}}

\usepackage{tikz,epstopdf}

\usetikzlibrary{arrows}
\usepackage{caption}
\usepackage{subfig}

\usepackage{lipsum,graphicx}
\usepackage[para]{threeparttable}
\usepackage{epsfig,graphics,graphicx,subfig,amssymb,amstext,amsmath,algorithm, algorithmic,wrapfig,multirow}
\usepackage{adjustbox}
\usepackage{siunitx}
\usepackage{pgfplots}
\usepackage{xcolor}
\usepackage{url}

\usepackage{breakurl}
\usepackage[breaklinks]{hyperref}
\usetikzlibrary{chains,arrows,calc ,positioning}
\usepackage{graphicx}
\usepackage{colortbl, xcolor}
\usepackage{enumitem}
\setitemize{noitemsep,topsep=0pt,parsep=0pt,partopsep=0pt}
\def\BibTeX{{\rm B\kern-.05em{\sc i\kern-.025em b}\kern-.08em T\kern-.1667em\lower.7ex\hbox{E}\kern-.125emX}}

\usepackage[left=1.62cm,right=1.62cm, top=0.75in, bottom = 1.05in]{geometry}

\begin{document}
\title{A Geometry-based Stochastic Wireless Channel Model using Channel Images}
\author{
\IEEEauthorblockN{Seongjoon Kang}
\IEEEauthorblockA{NYU Wireless, Brooklyn, NY, USA}
}


\maketitle
\begin{abstract}
Due to the high complexity of geometry-deterministic wireless channel modeling and the difficulty in its implementation, geometry-based stochastic channel modeling (GBSM) approaches have been used to evaluate wireless systems. 
This paper introduces a new method to model any GBSM by training a generative neural network using images formed by channel parameters.
In this work, we obtain channel parameters from the ray-tracing simulation in a specific area and process them in the form of images to train any generative model. 
Through a case study, we confirm that the use of channel images completes the training of the generative model within \textbf{$10$} epochs.  
In addition, we show that the trained generative model based on channel images faithfully represents the distributions of the original data under the assigned conditions. Therefore, we argue that the proposed data-to-image methods will facilitate the modeling of GBSM using any generative neural network under general wireless conditions. 
\end{abstract}


\section{Introduction}
The goal of communication channel modeling is to facilitate link- and system-level simulations in which the performance of the wireless system is evaluated. The $3$rd Generation Partnership Project ($3$GPP) proposes different types of stochastic channel model depending on wireless systems such as aerial and satellite communications for benchmarking \cite{3GPP38901, 3GPP36777, 3GPP38811}.

The statistical channel models proposed by $3$GPP aim at general scenarios such as urban micro- and rural regions and do not take into account geometry-specific scenarios. 
To consider geometry-specific scenarios, deterministic models such as using ray-tracing simulation are proposed on the basis of fundamental electromagnetic wave propagation theory. However, due to the higher running and implementation cost of the deterministic equivalent, geometry-based stochastic channel modeling (GBSM) is proposed \cite{yin2016propagation}, which models the channel parameters through statistical distributions of predefined local scatters in certain environments.

To simplify GBSM, data-driven generative neural networks are used \cite{yang2019generative, orekondy2022mimo, xia2022generative, hu2022multi, xiao2022channelgan, seyedsalehi2019propagation}. However, prior efforts to model GBSM require the high complexities of implementation and do not generalize the modeling for any wireless condition. 
For example, the authors of \cite{xia2022generative} propose a variational autoencoder (VAE) to capture the channel parameter statistics. This work requires two-step procedures---link classifier and VAE---to generate channel parameters. Following the same implementation procedures, in \cite{hu2022multi}, the cluster generative model is proposed using Wasserstein generative adversarial networks with gradient penalty (WGAN-GP). However, this effort has pronounced shortcomings in the aspect of simulation accuracy, as more multipath components need to be observed in a wideband due to the higher time resolution.

Therefore, it is desirable to simplify and generalize the implementation of GBSM using any generative model to capture the statistics of the parameters of multipath channels, such as path loss, propagation delay, and arrival and departure angles.  
In this work, we propose methods to convert the parameters of wireless channels to certain types of images, which we define as \emph{ channel images}, and argue that newly generated channel images facilitate training of the given generative model and reduce the complexity of implementation.  In addition, through a case study, we show that the output values of the trained model are statistically well matched to those from the original data.
\section{Channel Image Generation}
In this section, we discuss the proposed methods for processing and creating images from raw data on channel parameters before training a generative model. Since measuring realistic wireless channels in any region is challenging, we run the ray-tracing simulator and obtain the data from it.

\label{sec:data_processing}

\subsection{Channel Parameters from Ray-tracing Simulation}
The parameters of the wireless channel for a specific region can be obtained from the commercial ray tracer, Wireless Insite\cite{remcomm}. For each wireless link, we consider all possible propagation path parameters which are the resultant outputs from ray tracing simulator. With those parameters, we can model any channel including multiple-input and mltiple-output (MIMO) channels.

Specifically, for each link $i$ between a transmitter (TX) and receiver (RX), the channel parameter matrix $\boldsymbol{D}^i \in \boldsymbol{R}^{8 \times N}$ obtained from the ray tracing simulator is represented below.
\begin{equation} \boldsymbol{D}^i =
\left (
\begin{array}{cccc}
{pl}_1 & {pl}_2 & \ldots & {pl}_N\\
{dly}_1 & {dly}_2 & \ldots & {dly}_N\\
{aod}_1 & {aod}_2 & \ldots & {aod}_N\\
{zod}_1 & {zod}_2 & \ldots & {zod}_N\\
{aoa}_1 & {aoa}_2 &  \ldots & {aoa}_N\\
{zoa}_1 & {zoa}_2 & \dots & {zoa}_N\\
{ps}_1 & {ps_2} &  \dots & {ps}_N\\
{ls}_1 & {ls}_1 & \ldots & {ls}_1 \\
\end{array}
\right )
\label{eq:data_mat}
\end{equation}
where $N$ is the total number of multipaths, $pl$ is pathloss, $dly$ is propagation delay,  $aod$ and $zod$ are azimuth and zenith angles of departure (AOD and ZOD), $aoa$ and $zoa$ are azimuth and zenith angles of arrival (AOA and ZOA), $ps$ is the arrival phase at receiver side, and $ls_1$ is the link state of the first arrival path. Note that we only need the link state of the first arrival path since the other paths are non-line-of-site (NLOS) paths. The single-matrix representation for all channel parameters on one link is useful in building channel images. 

For line-of-site (LOS) path components, which is the first column of $\boldsymbol{D}^i$, are determined mathematically as the followings:
\begin{subequations}
\vspace{-2mm}
\begin{align}
    &pl_1 = 20\log_{10}({\rm dist}_{\rm 3d})+20\log_{10}(f)-147.55 \label{eq:los_pl}\\
    &dly_1 = \frac{{\rm dist}_{3\rm d}}{c} \label{eq:dly}\\
    &aod_1 = \arctan{\frac{y_{\rm tx}-y_{\rm rx}}{x_{\rm tx} - x_{\rm rx}}} \label{eq:aod}\\
    &aoa_1  =  aod_1 - 180 \label{eq:aoa}\\
    &zod_1 = \arctan{\frac{{\rm dist}_{\rm 2d}}{z_{\rm rx}-z_{\rm tx}}} \label{eq:zod}\\
    &zoa_1 = 180 - zod_1 \label{eq:zoa}\\
    &ps_1 = -360 (f \cdot dly_1 - \left\lfloor f \cdot dly_1 \right\rfloor)
    \label{eq:los_ps}
\end{align}
\end{subequations}
where $\rm {dist}_{3d}$ is $3$D distance between a TX and a RX, $f$ is the carrier frequency, $c$ is the speed of light, ($x_{\rm tx}$, $y_{\rm tx}$) and ($x_{\rm rx}$, $y_{\rm rx}$) are the $2$D coordinates of the TX and the RX. 

\subsection{Data Normalization}
In order to use a generative neural network, the raw data obtained by the ray-tracing simulation are normalized by employing \emph{Min-Max} scaling to make all elements of the data matrix $\boldsymbol{D}^i \in \boldsymbol{R}$ in the range of -$1$ to $1$.
The pre-processing of raw data is a common method to reduce the variances.  

Suppose that the $M$ TX - RX links are simulated in ray-tracing simulation. Since each link has a different number of paths, the corresponding data matrices $\boldsymbol{D}$ have different data shapes. 
To make the size of the data matrix the same across all links, we introduce the \emph{virtual paths} when the number of paths is less than the maximum number of paths, which is $25$ in this work. Except the pathloss feature, the virtual values for the feature data $\boldsymbol{d}_i$ are sampled from the uniform distribution $U(\rm{min}~\boldsymbol{d}_i,\rm{max}~\boldsymbol{d}_i)$.
 For the pathloss feature, we choose virtual values that are higher than the maximum pathloss value---\SI{180}{dB}, since the pathloss values will become the main criteria for distinguishing virtual and real paths.
In this way, we make the dimension of every data matrix $\boldsymbol{D}^i$ the same and denote the total data tensor $\mathcal{\boldsymbol{D}}^{\rm total} = \{\boldsymbol{D}^1, \boldsymbol{D}^2, \ldots, \boldsymbol{D}^M \}\in \boldsymbol{R}^{M \times 8 \times 25}$. 

Before Min-Max scaling for the raw data, we take into account the following data processing technique for some features.
\begin{itemize}
    \item {Pathloss}: Since we assume that the TX and RX locations are known, we normalize the pathloss values per path by the free space pathloss (FSPL) using Eq.~\ref{eq:los_pl}
    \begin{equation*}
        \Tilde{\boldsymbol{pl}} = \boldsymbol{pl} - FSPL(\rm {dist}_{3d})
    \end{equation*}
    \item {Delay}: Similarly, the delay values are normalized by the LOS delay using Eq.~\ref{eq:dly} 
    \begin{equation*}
        \Tilde{\boldsymbol{dly}} = \boldsymbol{dly} - \frac{\rm {dist}_{3d}}{c}
    \end{equation*}
    Considering that delay values are relatively small compared to other feature data, we multiply them by a sufficiently large scalar value---$10^7$.
    \item{Link state}:  
     The link state values of the first arrival path are sampled from uniform distributions, which are $U (1-\epsilon, 1)$ for LOS and $U(-1, -1+\epsilon)$ for NLOS, respectively, for a very small value $\epsilon$ ($0<\epsilon\ll1$),
\end{itemize}

 We do not consider angles relative to the LOS directions because this processing might not be invertible under certain conditions.
 For example, when the conditional variables are the 2D distance between a TX and RX and the height of a RX, the AOD and AOA are not recoverable from angles relative to LOS directions.

The Min-Max scaling for each feature data is given as the following:
\begin{align*}
    \hat{\boldsymbol{d}_i} &= \frac{(2 \times  \boldsymbol{d}_i - \max{\boldsymbol{d}_i} - \min{\boldsymbol{d}_i})}{\max{\boldsymbol{d}_i} - \min{\boldsymbol{d}_i}}\\
    \boldsymbol{d}_i &= \mathcal{\boldsymbol{D}}^{\rm total}_{:,i,:}
\end{align*}
Note that all the pre-process for raw data must be invertible, since we need the inverse process to recover data from the output of the generative model. 

\begin{figure}[!b]
\centering
\subfloat[][$25$ paths]
{\includegraphics[width =0.45\columnwidth]{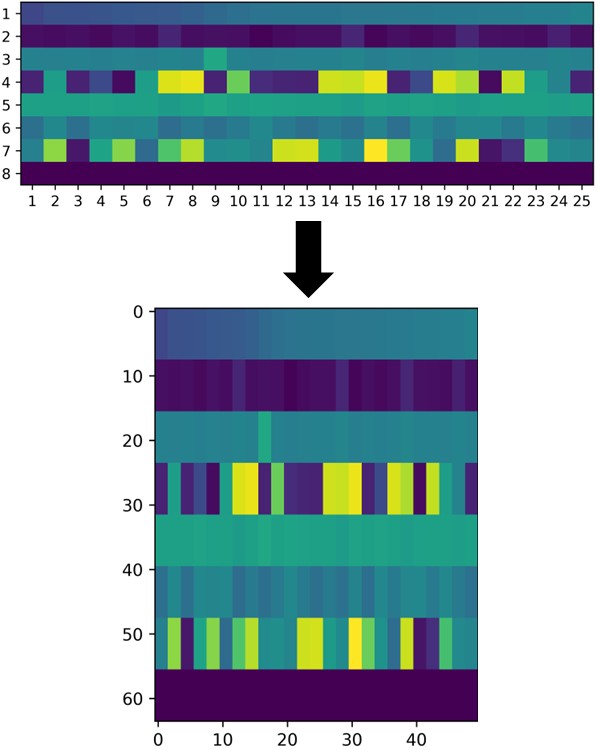} \label{fig:25paths}}
\subfloat[][$13$ paths]
{\includegraphics[width =0.45\columnwidth]{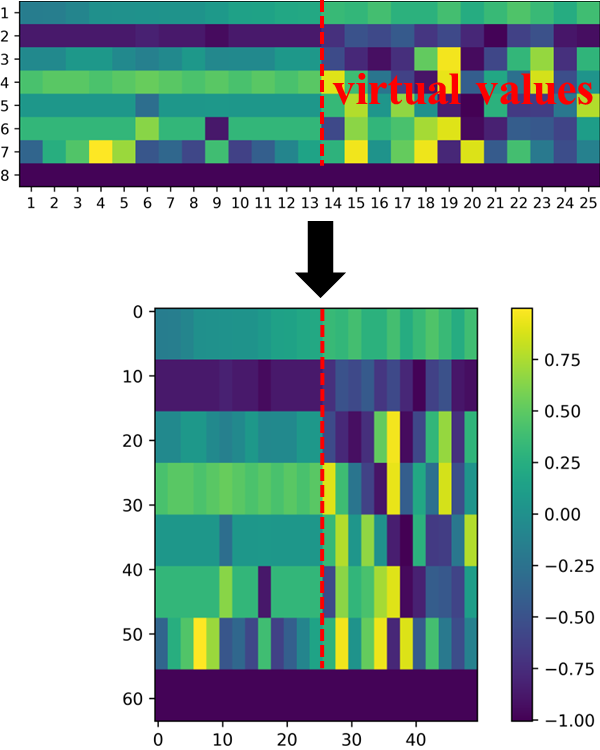} \label{fig:13paths}}
\caption{Images for different number of multipaths are generated by duplicating each element of the pre-processed channel matrix horizontally and vertically.}
\label{fig:images_from_two}
\end{figure}
\subsection{Data-to-Image Mapping}
After pre-processing the data, we create channel images for each wireless link between a TX and RX. The formation of images is instrumental to capture \emph{correlations} between multipath components and facilitate the training because deep convolutional generative networks can easily detect edges and capture correlations between pixels. 
In this work, we simply enlarge the data matrix $\boldsymbol{D}^i$ by replicating each element of the matrix horizontally and vertically. 
Specifically, we copy each element of the data matrix $\boldsymbol{D}^i \in \mathbb{R}^{8 \times 25}$ by $2$ times horizontally and $8$ times vertically, so that the size of the resulting matrix becomes $\Tilde{\boldsymbol{D}}^i \in \mathbb{R}^{64 \times 50}$. 
\begin{equation*}
    \boldsymbol{D}^i \in \mathbb{R}^{8 \times 25} \rightarrow \Tilde{\boldsymbol{D}}^i_{\rm image} \in \mathbb{R}^{64 \times 50}
\end{equation*}

Fig.~\ref{fig:images_from_two} depicts the process of creating images from the original data matrices for two different multipath channel cases. 
As shown in Fig.~\ref{fig:13paths}, when the number of multipaths is $13$ ($<25$), the virtual values are filled to make the size of the data matrix consistent. 


\section{Design of Conditional Generative Model}
The main goal of GBSM in this work is to capture channel statistics for a specific region under certain conditions, such as RX heights, distances between TX and RX, and carrier frequencies. Therefore, we need to design a generative model that can conditionally output the channel parameters.
\begin{figure}[!h]
\centering
\includegraphics[width =0.99\columnwidth]{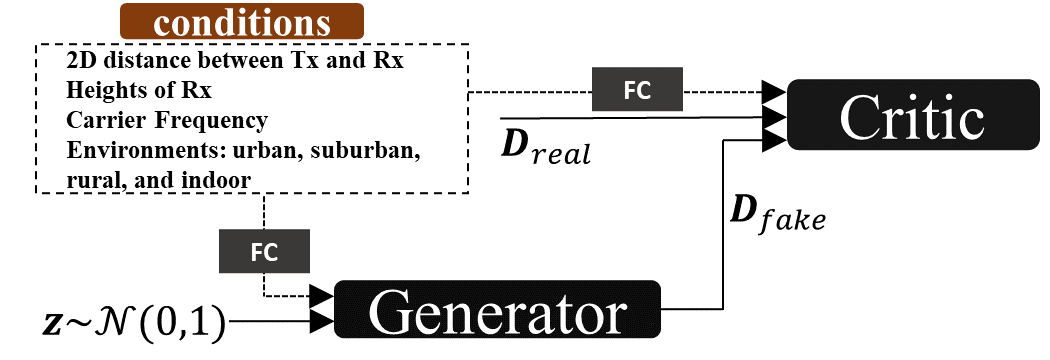} 
\caption{WGAN-GP with general wireless conditions }
\label{fig:w_gan}
\end{figure}

As an example of the conditional generative model,  we employ WGAN-GP \cite{gulrajani2017improved}.  For more detailed descriptions of WGAN and WGAN-GP, we refer to \cite{arjovsky2017wasserstein, gulrajani2017improved}. 
 Fig.~\ref{fig:w_gan} describes the structure of WGAN-GP with general wireless conditions, such as 2D distances, heights, carrier frequencies, and network environments, which can be embedded both of the generator and the critic, respectively, through fully connected neural networks (FC). 
The hyper-parameters for training are listed in Table~\ref{tab:gan_hyperparams}, and the structure of our WGAN-GP is similar to \cite{gulrajani2017improved}.  
All detailed implementations are uploaded in \cite{chanmod-github}.

\begin{table}[!h]
\begin{center}
\caption{Hyper-parameters to train WGAN-GP}
\label{tab:gan_hyperparams}
\begin{tabular}{|c|c|c|}
\cline { 2 - 3 } \multicolumn{1}{c|}{} & \begin{tabular}{c} 
 \textbf{Generator} \\
\end{tabular} & \begin{tabular}{c} 
 \textbf{Critic} \\
\end{tabular} \\
\hline Input image size & \multicolumn{2}{|c|}{[64, 50 ,1]} \\
\hline Learning rate & \multicolumn{2}{|c|} {$ 10^{-4}$} \\
\hline Optimizer & \multicolumn{2}{|c|}{ Adam ($\beta_1 = 0.5$, $\beta_2 = 0.9$)} \\
\hline Epochs & \multicolumn{2}{|c|}{10} \\
\hline Batch size & \multicolumn{2}{|c|}{256} \\
\hline Number of parameters & 4,074,726 & 4,016,597\\
\hline
\end{tabular}
\end{center}
\end{table}

 \section{Case Study and Ray-Tracing Simulation}
 In this section, we discuss the case study on GCSM with conditional constraints and explain the ray-tracing simulation to obtain the wireless channel parameters in the given region. Note that the main assumption throughout this work is that the TX and RX locations are known. 
\subsection{Case Study: Different RX Heights } 
AS a case study, we want to capture the statistical distribution of local scatters at different 2D distances between TXs and RXs. In addition, we take into account different RX heights with fixed carrier frequency. Thus, the learning process of the statistical distribution of the data matrix $\boldsymbol{D}$ can be described as follows:

\begin{align}
 P(\boldsymbol{D}|\text{dist2d}, \text{h} ) =  P (T{\mathcal{G}}( \text{dist2d}, \text{h}))
 \label{eq:cond_h}
\end{align}
where $\mathcal{G}$ is any conditional generative model---WGAN-GP in our work, and $T$ is an operator to perform image-to-data mapping process, i.e., the process from outputs of the model to the original data matrix $\boldsymbol{D}$. Note that we can add more conditional variables with the functions of 2D distance dist2d and height h, e.g., 3D distances and zenith angles.   
In this study, the five discrete RX heights, $\SI{1.6}{m}$, $\SI{30}{m}$, $\SI{60}{m}$, and $\SI{120}{m}$, are considered to model terrestrial and aerial channels simultaneously. Note that $2$D distance values obtained from all pairs of TX and RX are continuous in the range of the network area.

\subsection{Ray-tracing Simulation} 
To obatain channel parameters for the case study mentioned above, we run a ray-tracing simulation in Herald Square of New York City, which can be considered a typical dense urban area.
We select the upper midband carrier frequency $\SI{12}{GHz}$, which is recognized as the promising spectrum for future networks \cite{fcc2023preliminary, fcc20236Gworkinggroup}, and all electrical properties, such as permeability and conductivity, corresponding to $\SI{12}{GHz}$, are set for each material. 
The data from this ray-tracing simulation can be used to model integrated aerial-terrestrial wireless channels.

\begin{figure}[!b]
\centering
\includegraphics[width =0.97\columnwidth]{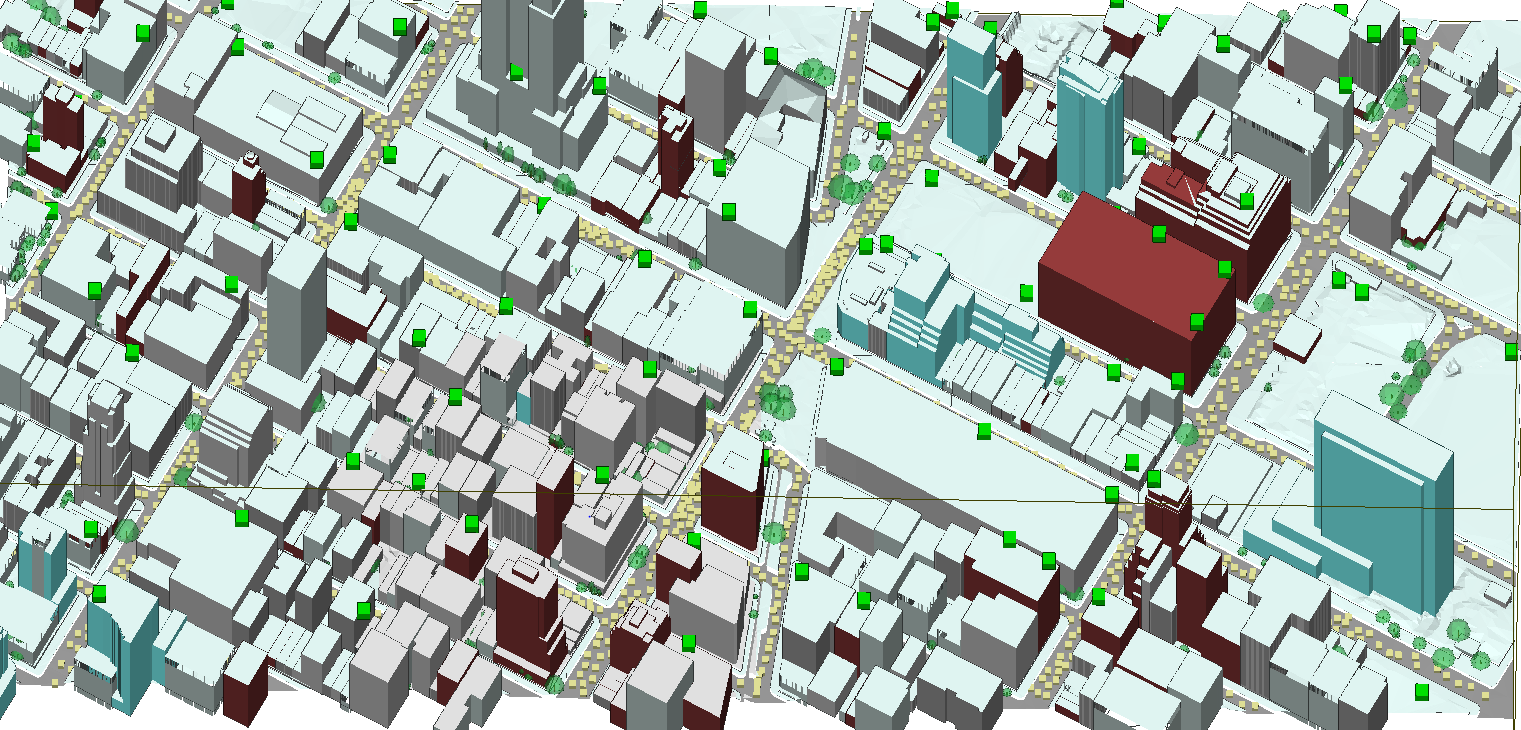} 
\caption{Herald Square in New York city for ray-tracing simulation. ($\SI{1120}{m} \times \SI{510}{m}$). The green rectangles represent TXs, and the yellow rectangles are RXs. }
\label{fig:boston}
\end{figure}

\begin{figure*}[!t]
\centering
\subfloat[][]
{\includegraphics[width =0.42\columnwidth]{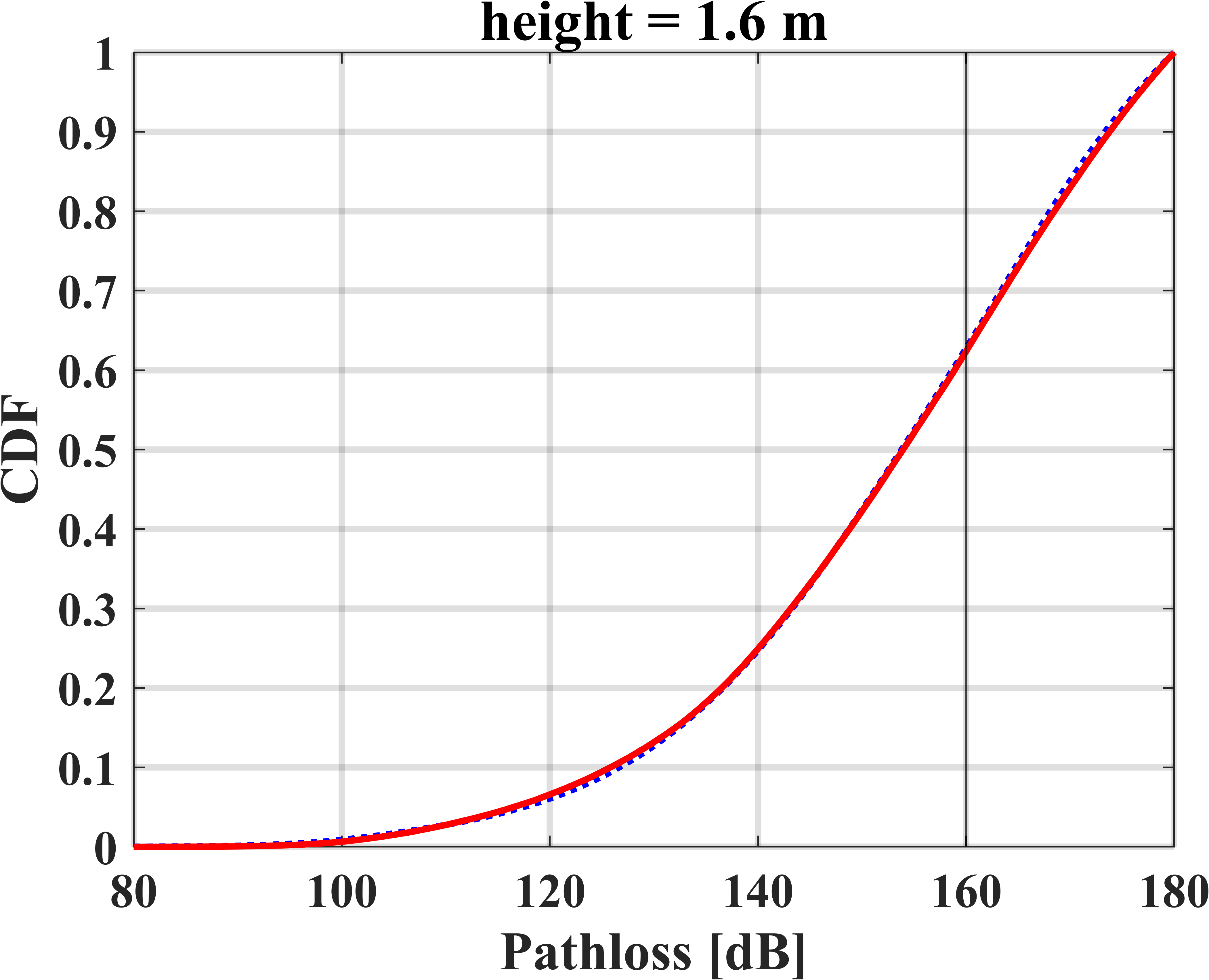} \label{fig:pl_1_6}}
\subfloat[][]
{\includegraphics[width =0.38\columnwidth]{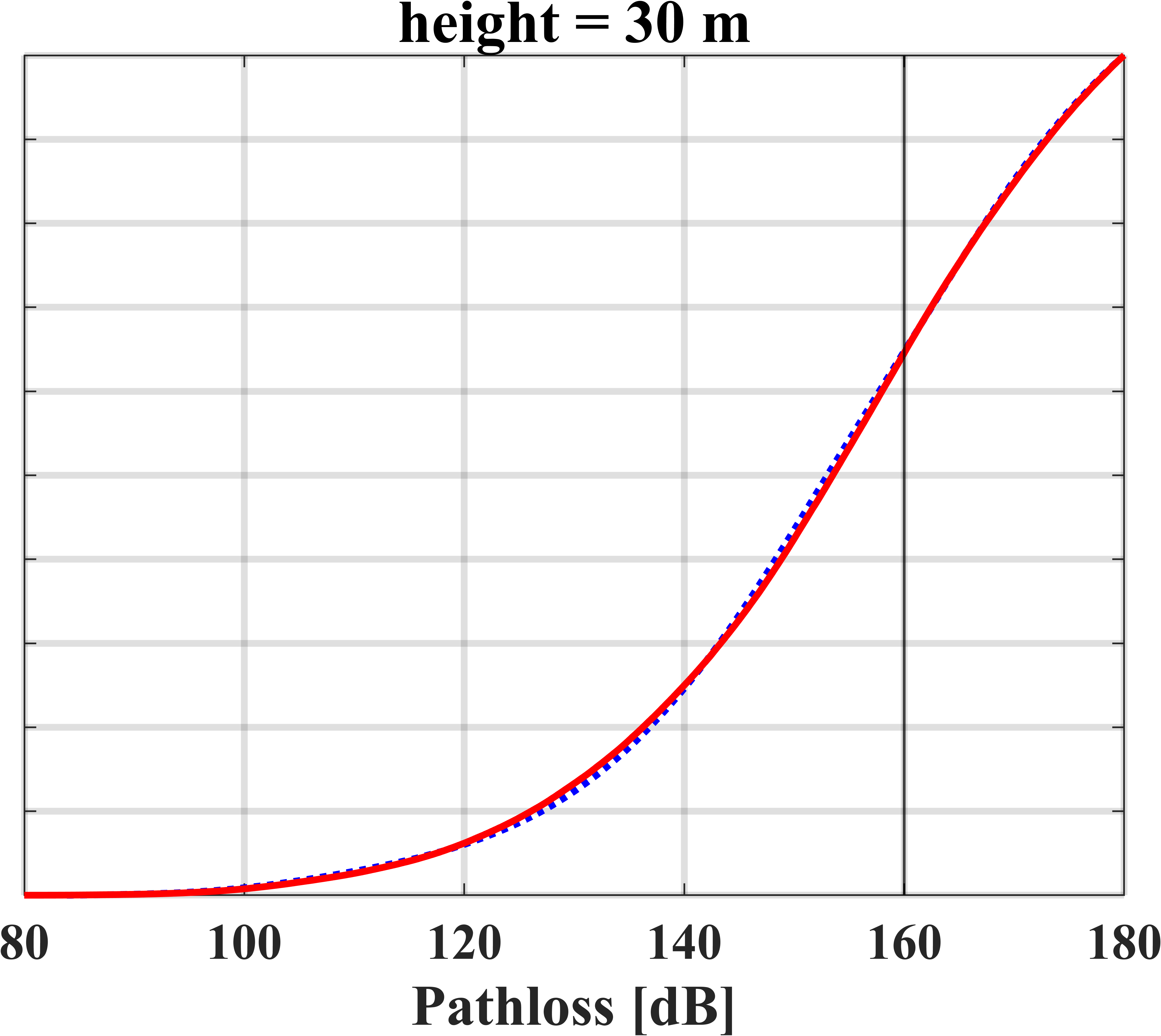} \label{fig:pl_30}}
\subfloat[][]
{\includegraphics[width =0.38\columnwidth]{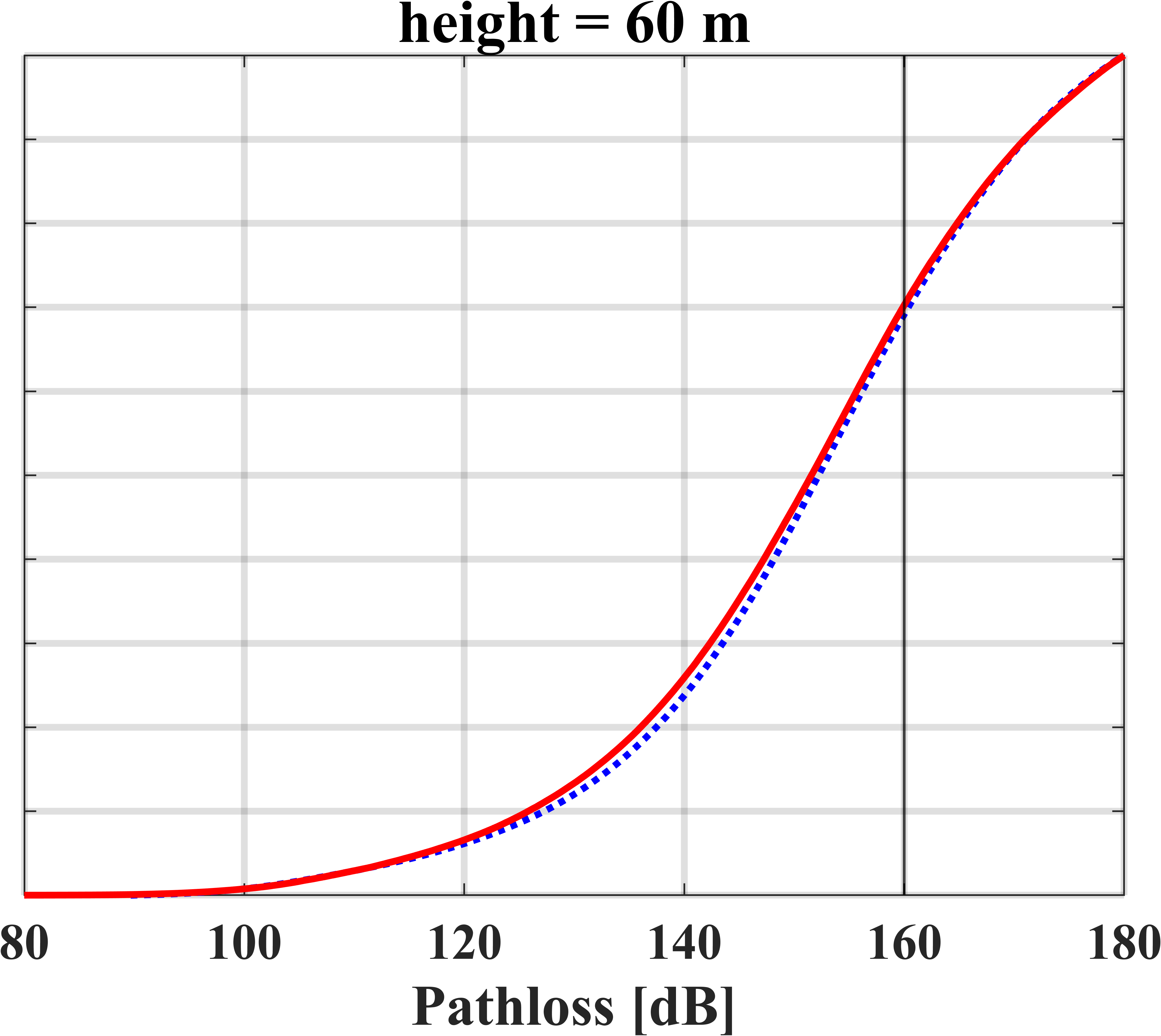} \label{fig:pl_60}}
\subfloat[][]
{\includegraphics[width =0.38\columnwidth]{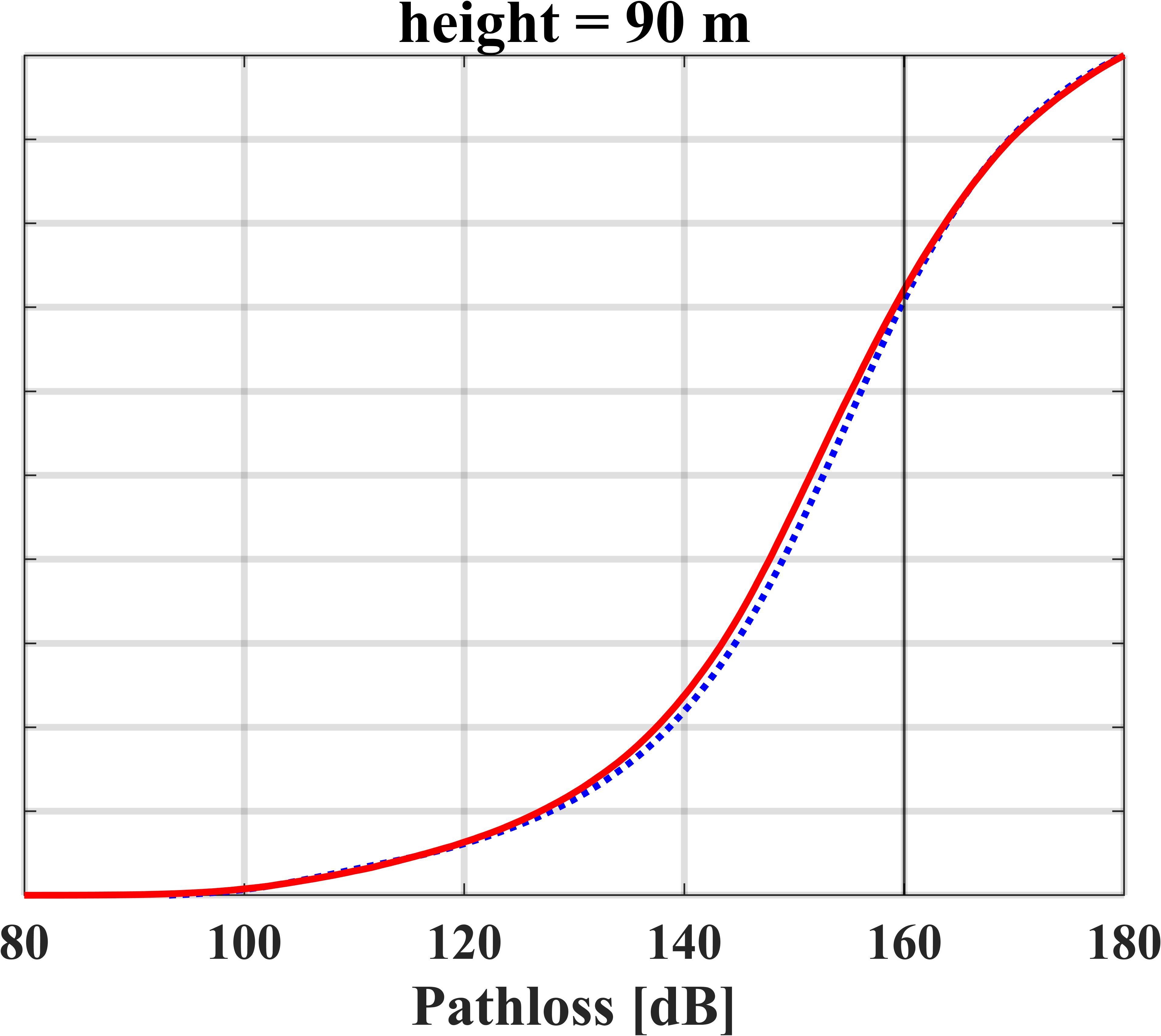} \label{fig:pl_90}}
\subfloat[][]
{\includegraphics[width =0.38\columnwidth]{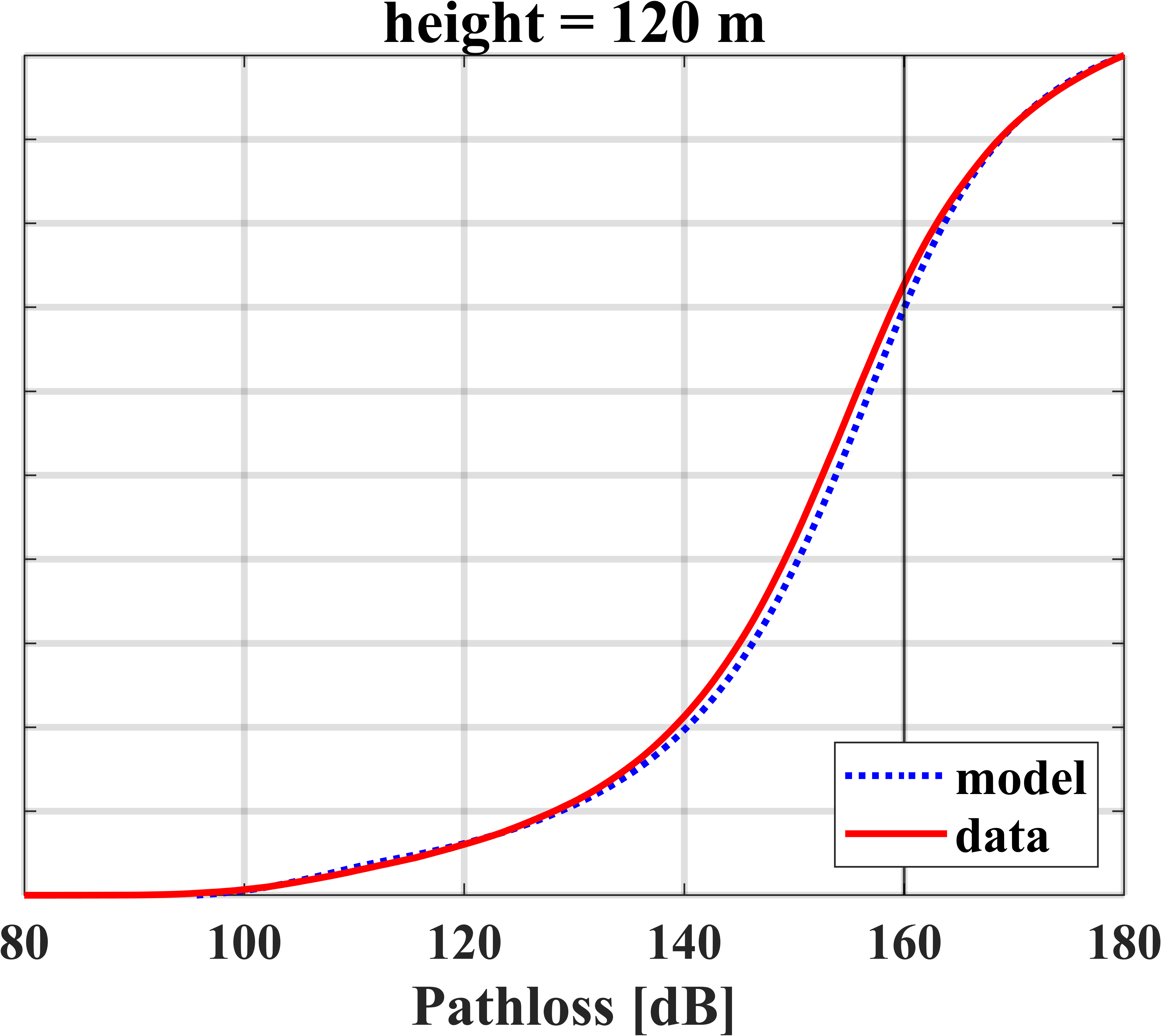} \label{fig:pl_120}}
\caption{CDFs of pathloss for (a) $\SI{1.6}{m}$, (b) $\SI{30}{m}$, (c) $\SI{60}{m}$, (d) $\SI{90}{m}$, and (e) $\SI{120}{m}$}
\label{fig:pl_cdf}
\end{figure*}

\begin{figure*}[!h]
\centering
\subfloat[][]
{\includegraphics[width =0.425\columnwidth]{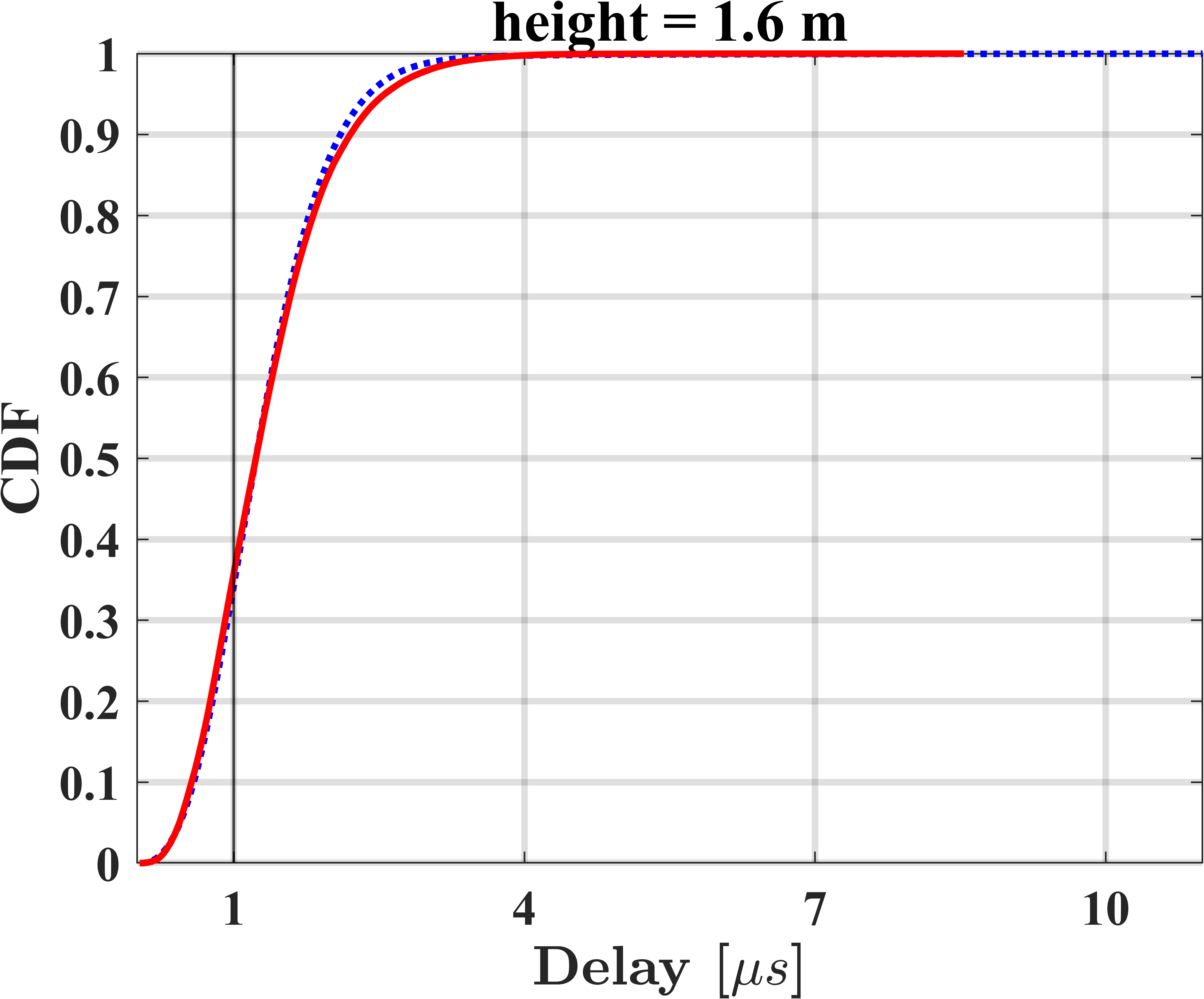} \label{fig:dly_1_6}}
\subfloat[][]
{\includegraphics[width =0.38\columnwidth]{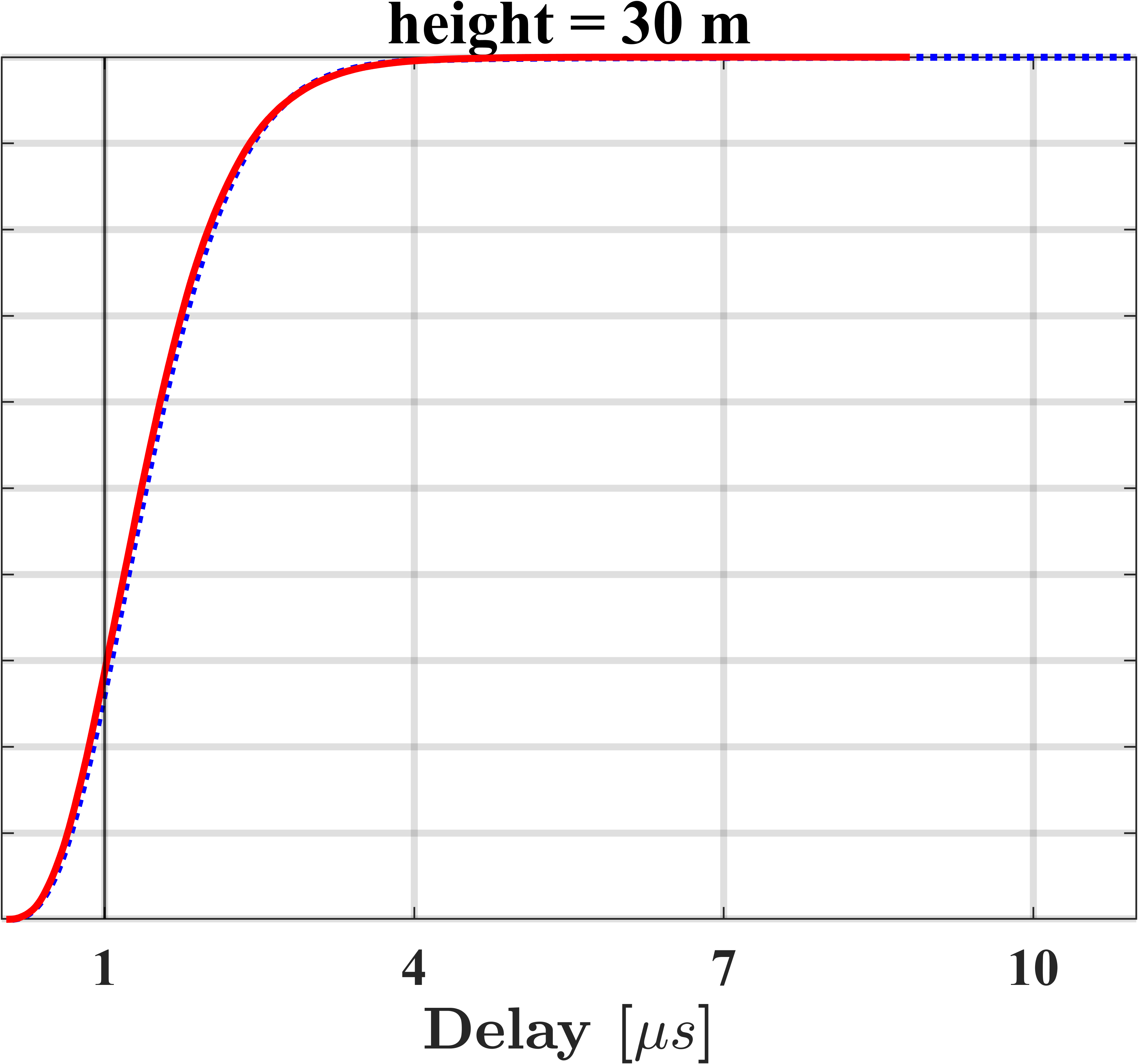} \label{fig:dly_30}}
\subfloat[][]
{\includegraphics[width =0.38\columnwidth]{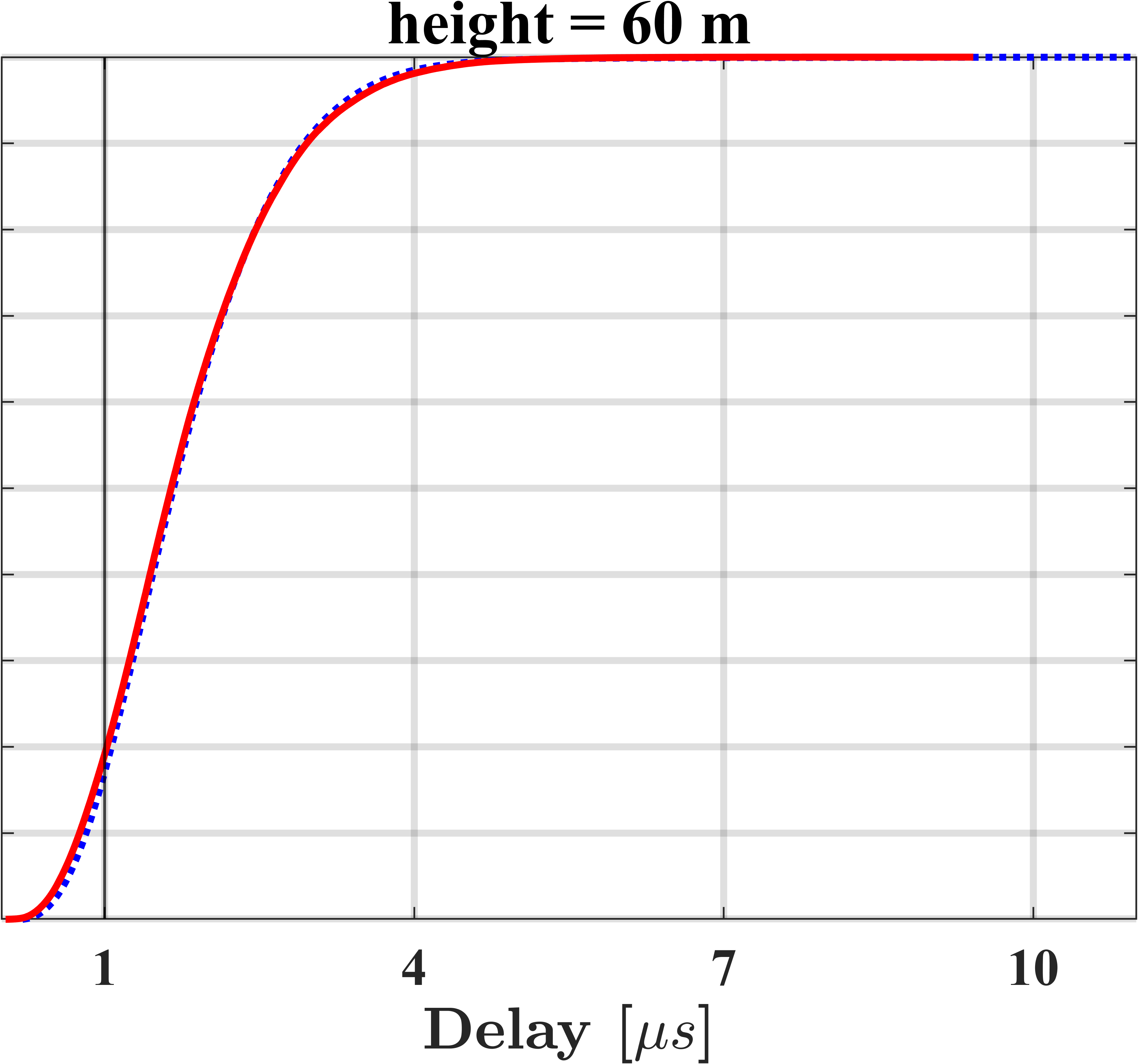} \label{fig:dly_60}}
\subfloat[][]
{\includegraphics[width =0.38\columnwidth]{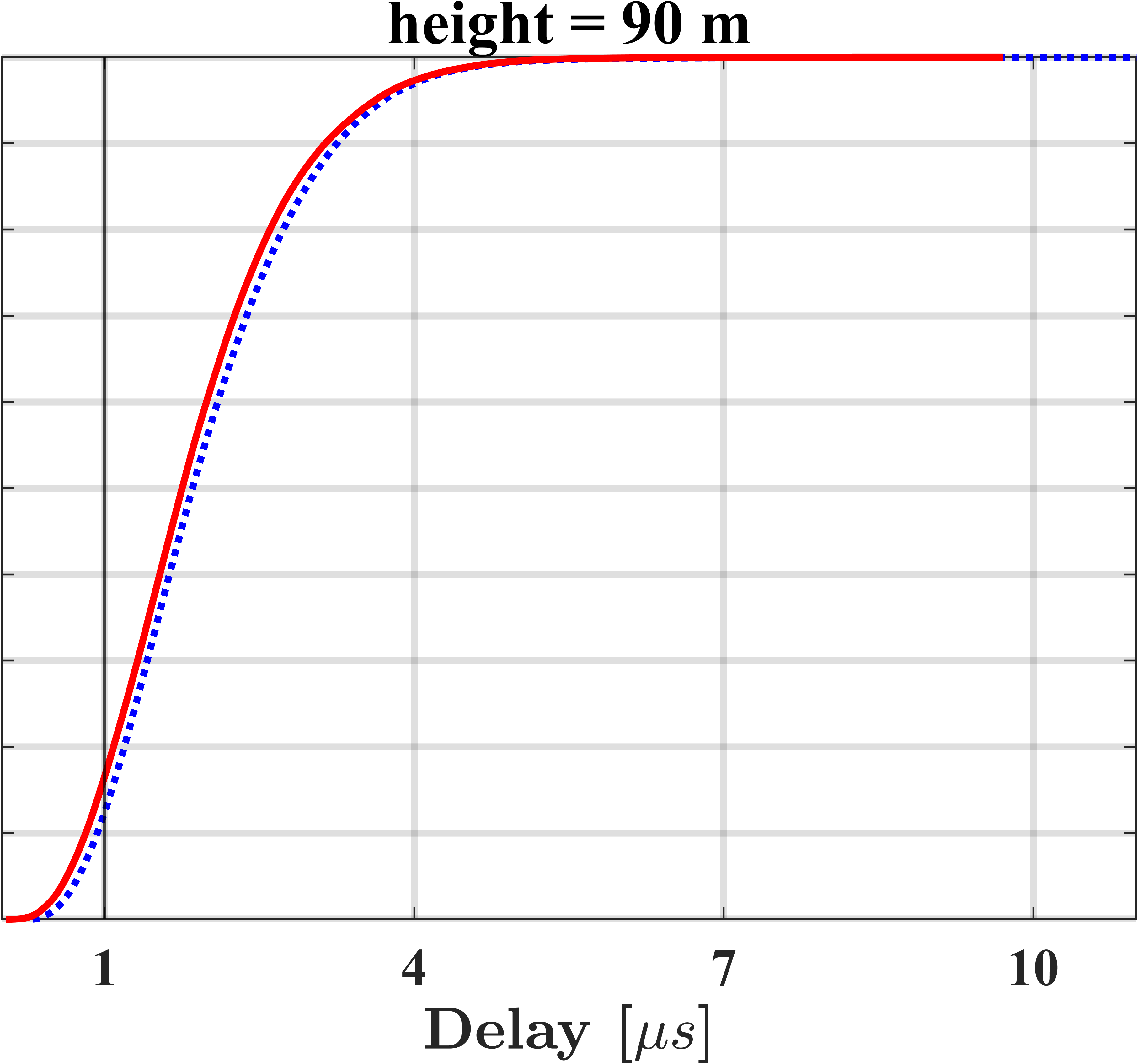} \label{fig:dly_90}}
\subfloat[][]
{\includegraphics[width =0.38\columnwidth]{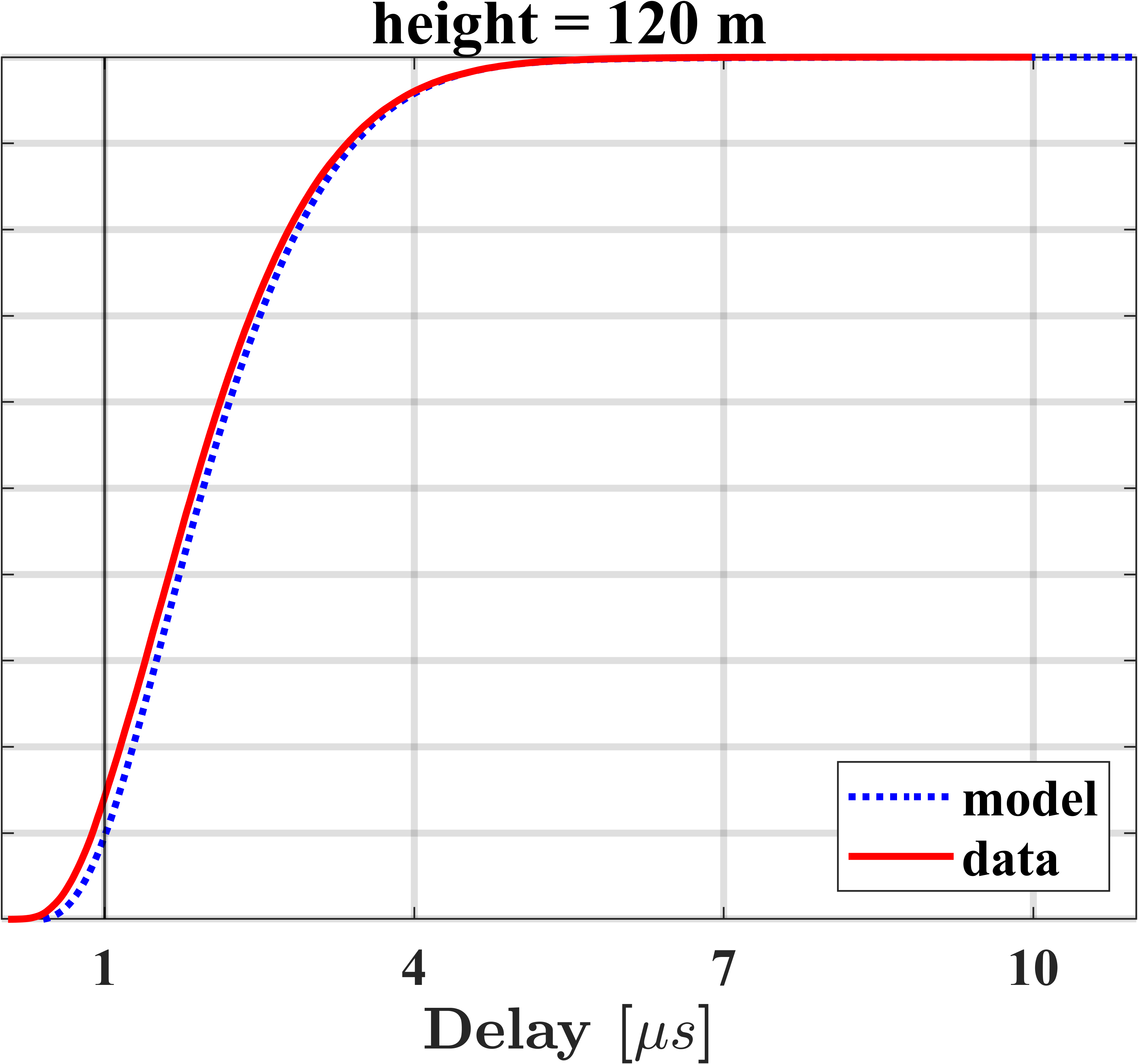} \label{fig:dly_120}}
\caption{ CDFs of propagation delay for (a) $\SI{1.6}{m}$, (b) $\SI{30}{m}$, (c) $\SI{60}{m}$, (d) $\SI{90}{m}$, and (e) $\SI{120}{m}$}
\label{fig:dly_cdf}
\end{figure*}

Fig.~\ref{fig:boston} shows the area, Herald Square in New York City, for the ray-tracing simulation where TXs are placed on top of buildings and RXs are dropped on the outdoor street randomly. 
In Fig.~\ref{fig:boston}, different colors indicate different materials. For example, gray is concrete, light blue is glass, and brown is brick. To capture transmitted rays in all directions, TXs and RXs are equipped with the single \emph{isotropic} antennas. 
In this simulation, we deploy $79$ TXs on top of buildings and $1526$ outdoor RXs for five different heights: 1.6, 30, 60, 90, $\SI{120}{m}$. Thus, in total $79\times5\times1526 = 602,770$ links are obtained. 

\section{Data recovering process}
We confirmed that within $10$ epochs of training, the channel images generated from the model become similar to the original. 
After the train, the output matrices $\Tilde{\boldsymbol{D}}_{\rm image} \in \boldsymbol{R}^{64 \times 50}$ from WGAN-GP are processed, reversely following all the steps described in Section \ref{sec:data_processing}.

\begin{itemize}
    \item First, the size of the output matrices should be decreased using a suitable sampling method: $\Tilde{\boldsymbol{D}}_{\rm image} \in \boldsymbol{R}^{64 \times 50} \rightarrow$ $\Tilde{\boldsymbol{D}} \in \boldsymbol{R}^{8 \times 25}$. 
    \item After reducing the size of the model output, all data are recovered by the \emph{reverse} Min-Max scaling. In particular, to recover pathloss and delay values, FSPL and minimum delay values calculated using Eqs.~\ref{eq:los_pl} and ~\ref{eq:dly} are added to pathloss and delay values after reverse Min-Max scaling.
    \item To determine the state of the link, we take the average in the last row of the output matrix $\Tilde{\boldsymbol{D}}$. If the average value is positive, the link state is LOS and, otherwise, NLOS. 
    \item Once a link state is determined as LOS, all channel parameters corresponding to the first arrival path are deterministic. 
Thus, when the link state value from $\Tilde{\boldsymbol{D}}$ indicates the LOS state, the first column of the data matrix $\Tilde{\boldsymbol{D}}$ is replaced with the channel parameters calculated by Eqs.~\ref{eq:los_pl}-~\ref{eq:los_ps} using the TX and RX coordinates. 
    \item Finally, virtual paths are discerned and removed when pathloss values are larger than the outage value ($\SI{180}{dB}$).

\end{itemize}

\section{Evaluation on Trained Model}

To evaluate the quality of the channel statistics captured by WGAN-GP, we analyze the output statistics from our model by comparing them with the original data.  


Fig.~\ref{fig:pl_cdf} shows the cumulative distribution functions (CDF) of pathloss on all paths over all links. We observe that the CDFs drawn by the data from the model are aligned with the CDFs by the original data. In particular, the conditionality by heights holds, as we confirm that the probability of $pl>\SI{160}{dB}$ decreases at higher altitudes in Fig.~\ref{fig:pl_cdf}. 
Similarly, as shown in Fig.~\ref{fig:dly_cdf}, we observe that two CDFs (model and data) of propagation delays on all paths of all links are well matched. In addition, at each height, we see the different conditional distributions, since the probability of $dly<1 \mu s$ decreases at higher altitudes.
This is mainly because in dense urban areas, the NLOS propagation paths are dominant when the RX heights are lower.

\begin{figure*}[!h]
\centering
\subfloat[][]
{\includegraphics[width =0.42\columnwidth]{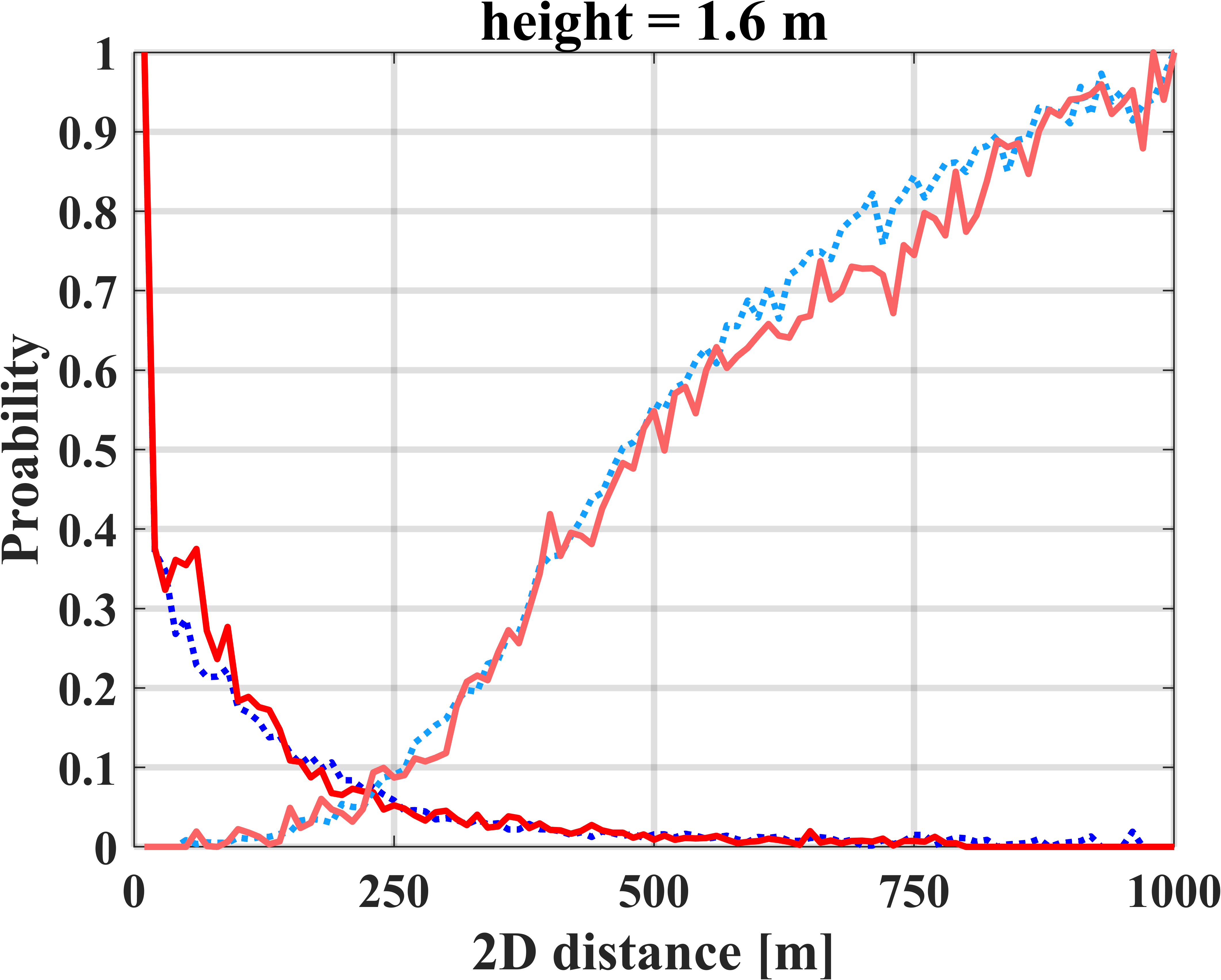} \label{fig:los_prob_1_6}}
\subfloat[][]
{\includegraphics[width =0.38\columnwidth]{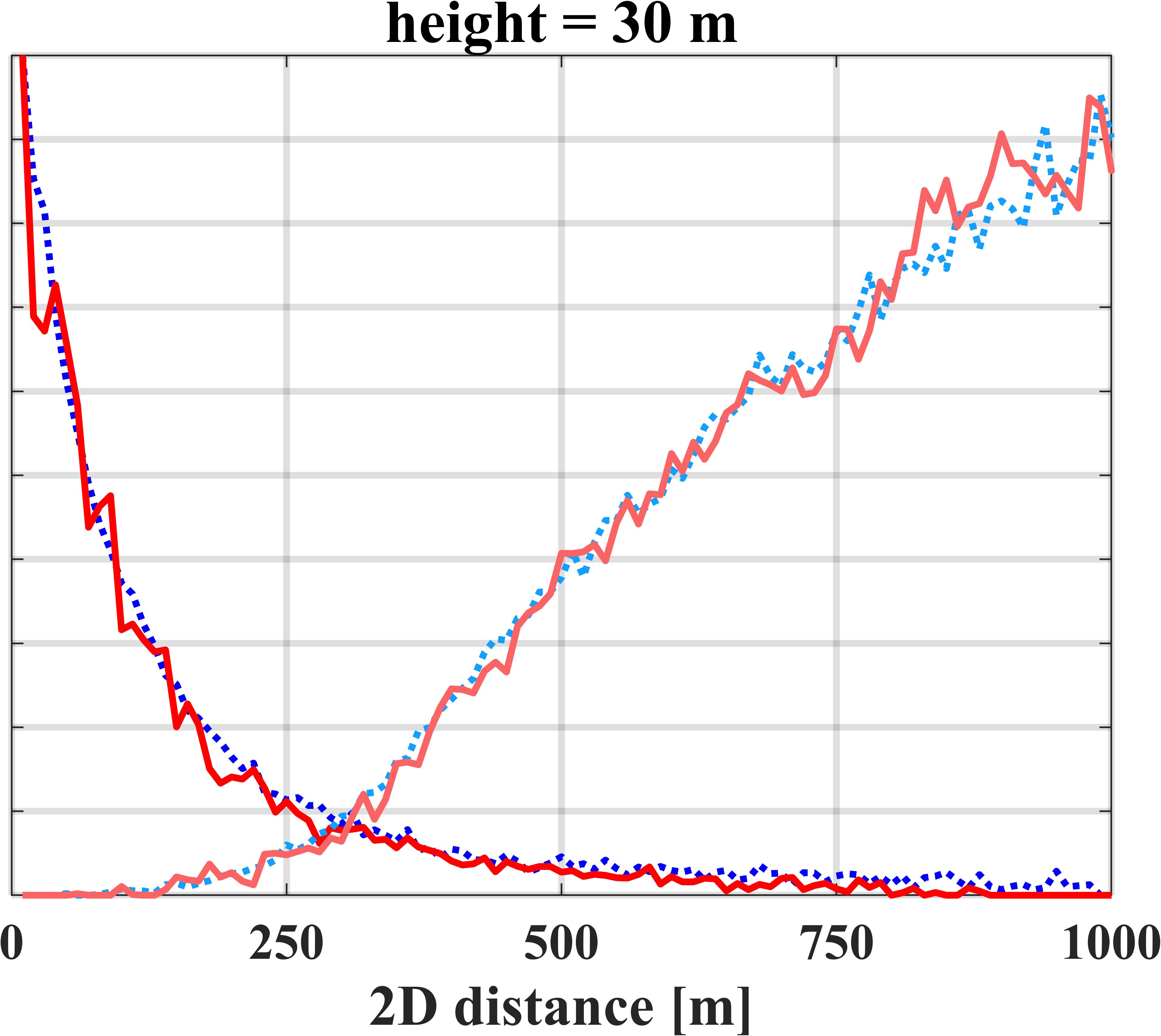} \label{fig:los_prob_30}}
\subfloat[][]
{\includegraphics[width =0.38\columnwidth]{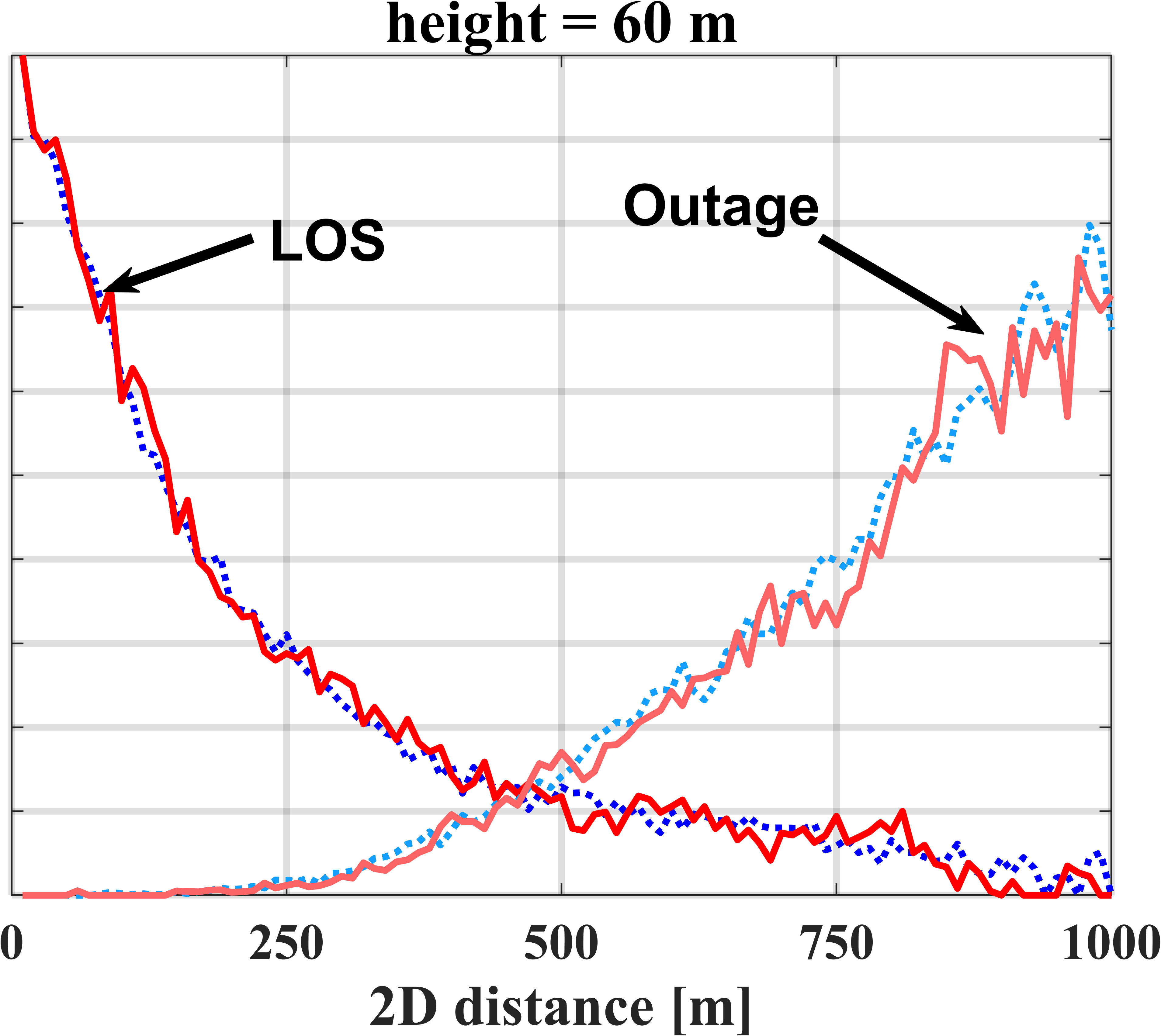} \label{fig:los_prob_60}}
\subfloat[][]
{\includegraphics[width =0.38\columnwidth]{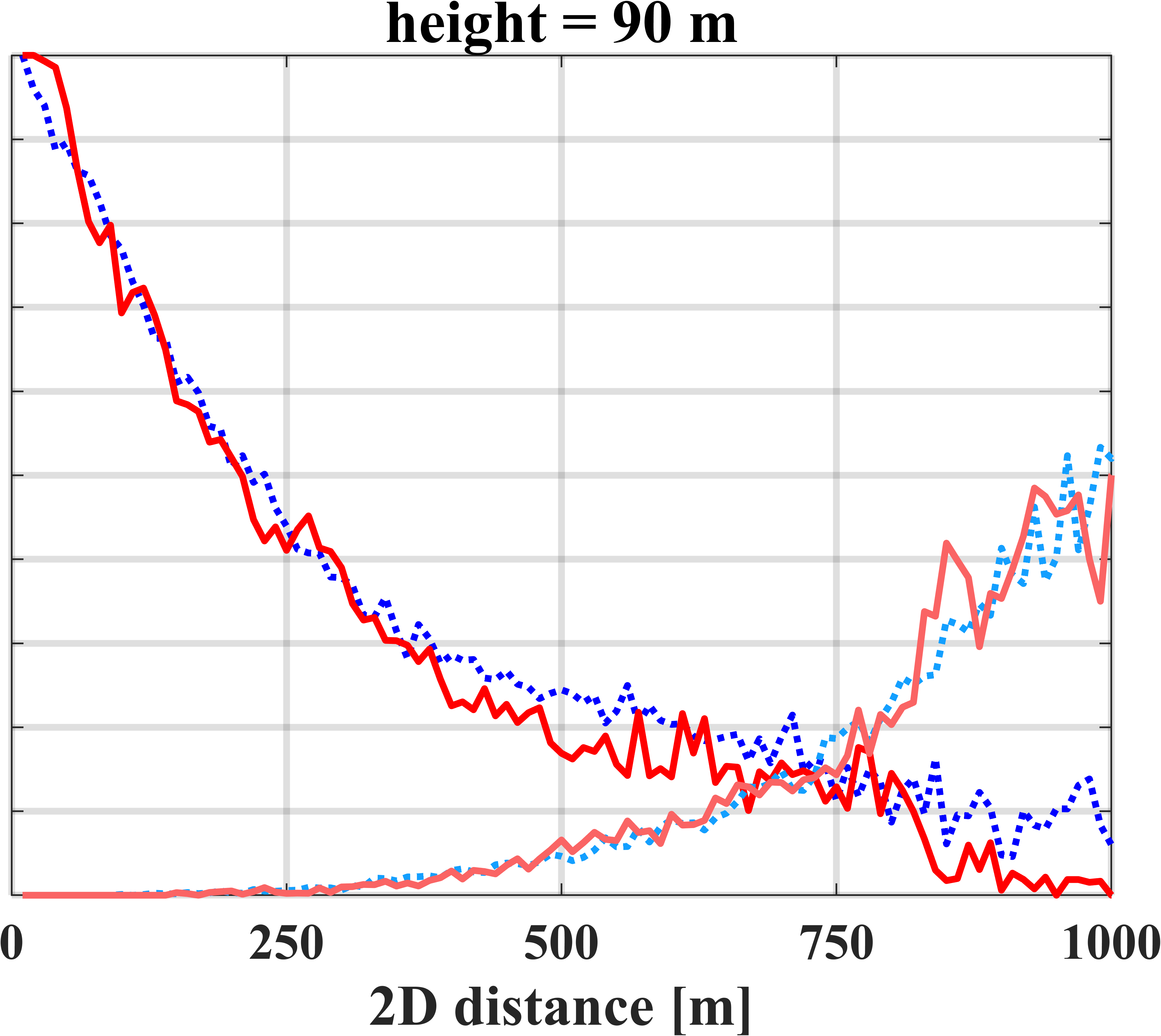} \label{fig:los_prob_90}}
\subfloat[][]
{\includegraphics[width =0.38\columnwidth]{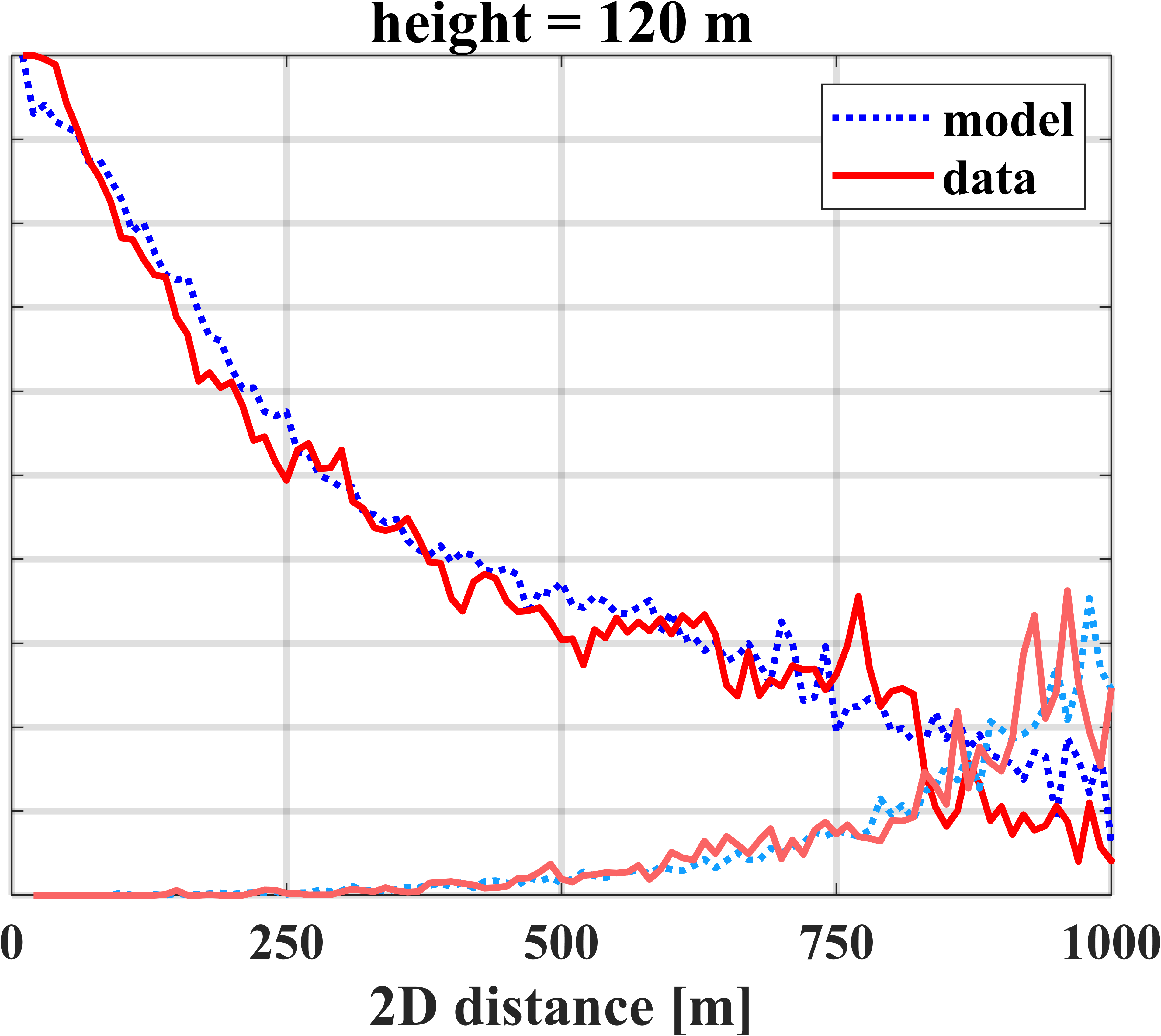} \label{fig:los_prob_120}}
\caption{ Link state probability (LOS and outage) for (a) $\SI{1.6}{m}$, (b) $\SI{30}{m}$, (c) $\SI{60}{m}$, (d) $\SI{90}{m}$, and (e) $\SI{120}{m}$}
\label{fig:los_cdf}
\end{figure*}

Fig.~\ref{fig:los_cdf} shows the probabilities of LOS and link failure at different 2D distances. We define a link outage if all pathloss values of a multipath channel are greater than $\SI{180}{dB}$.  At higher altitudes, the RXs are more visible to the TXs. Therefore, the probability of LOS is higher at higher altitudes, whereas the probability of outage is lower. We observe that the link state probabilities (LOS and outage) at each height obtained from our model are well matched with the probabilities calculated with the original data. This result implies that the trained model holds two conditional constraints well: 2D distances and heights.

To compare the angular distributions between our model and the original data, we took into account the angles relative to the LOS directions. In this case, we can only evaluate the statistics of the ZOD and ZOA under the conditions of 2D distances and heights. We define a random variable, a relative angle to the LOS direction, simply by subtracting the ZOD and ZOA of the LOS, which can be computed using Eq.~\ref{eq:zod} and ~\ref{eq:zoa}. Consequently, the probability density function (PDF) for the relative angle of ZOD conditioned with 2D distance ${\rm dist_{\rm 2D}}$ and height ${\rm h}$ is described as follows:
\begin{align}
    P( zod(&{\rm dist_{\rm 2D}}, {\rm h}) - {zod_{\rm LOS}({\rm dist_{\rm 2D}}, {\rm h})}) \nonumber \\
    &=  P\left( {zod}({\rm dist_{\rm 2D}}, {\rm h}) -  \arctan({\frac{{\rm dist}_{\rm 2D}}{\rm h}})\right)
    \label{eq:pdf_zoda}
\end{align}
In the same way, we define the PDF of ZOAs relative to LOS direction using Eq.~\ref{eq:zoa}.

\begin{figure}[!t]
\centering
\subfloat[][]
{\includegraphics[width =0.99\columnwidth]{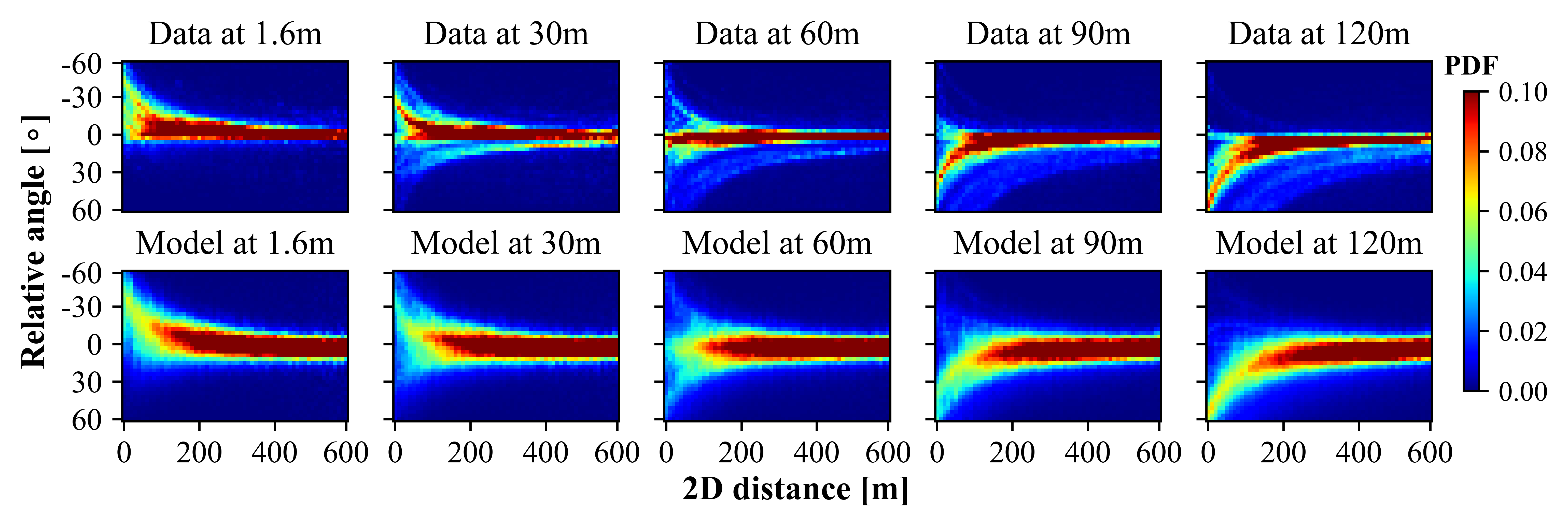} \label{fig:pdf_zod}}
\\
\subfloat[][]{
\includegraphics[width =0.99\columnwidth]{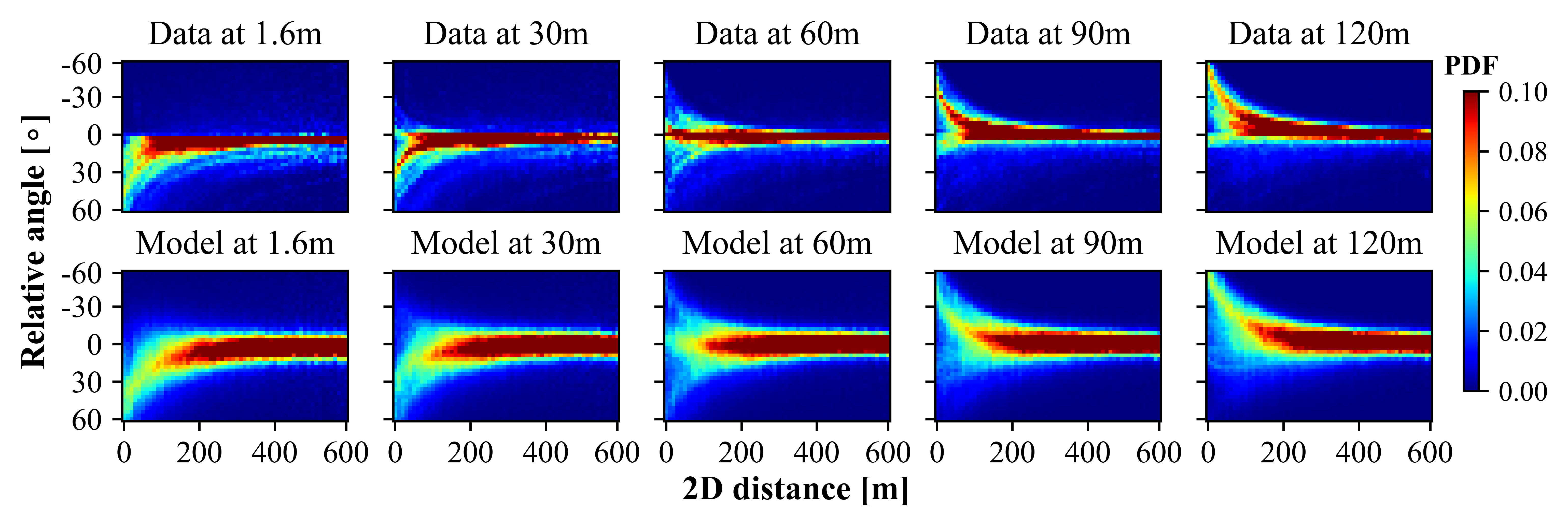}
\label{fig:pdf_zoa}}
\caption{ PDFs of relative (a) ZOD and (b) ZOA  over 2D distances}
\label{fig:pdf_zod_zoa}
\end{figure}

Fig.~\ref{fig:pdf_zod_zoa} depicts the PDFs of ZOA and ZOD as heatmaps obtained from the original data and the outputs of our model at different RX heights. Each column in the heatmaps represents the PDF calculated at the corresponding 2D distance. 
On the one hand, as the 2D distances increase, we observe that ZOA and ZOD have a less angular spread as a result of less scattering at a far distance. 
On the other hand, since the local scattering increases at a closer distance, we see a large variation of the relative angles until \SI{200}{m} of the 2D distance.

Specifically, local scatterings occur mostly in downward directions at close distances. Accordingly,  Fig.~\ref{fig:pdf_zod} and Fig.~\ref{fig:pdf_zoa} illustrate the following interesting results:
\begin{itemize}
    \item On the TX side, we can easily confirm that the LOS ZOD decreases as the RX height increases.   
As shown in Fig.~\ref{fig:pdf_zod}, at \SI{1.6}{m} altitude, we observe a higher probability at negative relative angles (LOS ZODs $>$ ZODs). On the contrary, at \SI{120}{m} altitude, positive relative angles (ZODs $>$ LOS ZODs) have a higher probability as the LOS ZODs decrease. 

\item On the RX side, as opposed to the TX case, the LOS ZOA increases as the RX height increases. 
Therefore, in Fig.~\ref{fig:pdf_zoa}, the positive relative angles (ZOAs $>$ LOS ZOAs) have higher probabilities at $\SI{1.6}{m}$ altitude, while at $\SI{120}{m}$ altitude, the probability of negative relative angles (LOS ZOAs $>$ ZOAs) is higher.
\end{itemize}

Since the conditional variables are 2D distances and heights, azimuth angles (AOA, AOD) and arrival phases follow the \emph{uniform} distribution.
Note that for AOA and AOD statistics, Eq.~\ref{eq:pdf_zoda} is not available and, therefore, it is not possible to show heatmaps of PDFs such as Fig.~\ref{fig:pdf_zod} and Fig.~\ref{fig:pdf_zoa}.
In other words, to calculate LOS azimuth angles, Eq.~\ref{eq:aod} is not available due to  
\begin{equation}
 P (aod |\text{dist}_{\text{2D}}, \text{h}) \neq   P ( aod |\text{dx}, \text{dy}, \text{h}) \nonumber
\end{equation}
where $\rm dx = x_{rx} - x_{tx}$ and $\rm dy = y_{rx} - y_{tx}$.
\begin{figure}[!t]
\centering
\includegraphics[width =0.99\columnwidth]{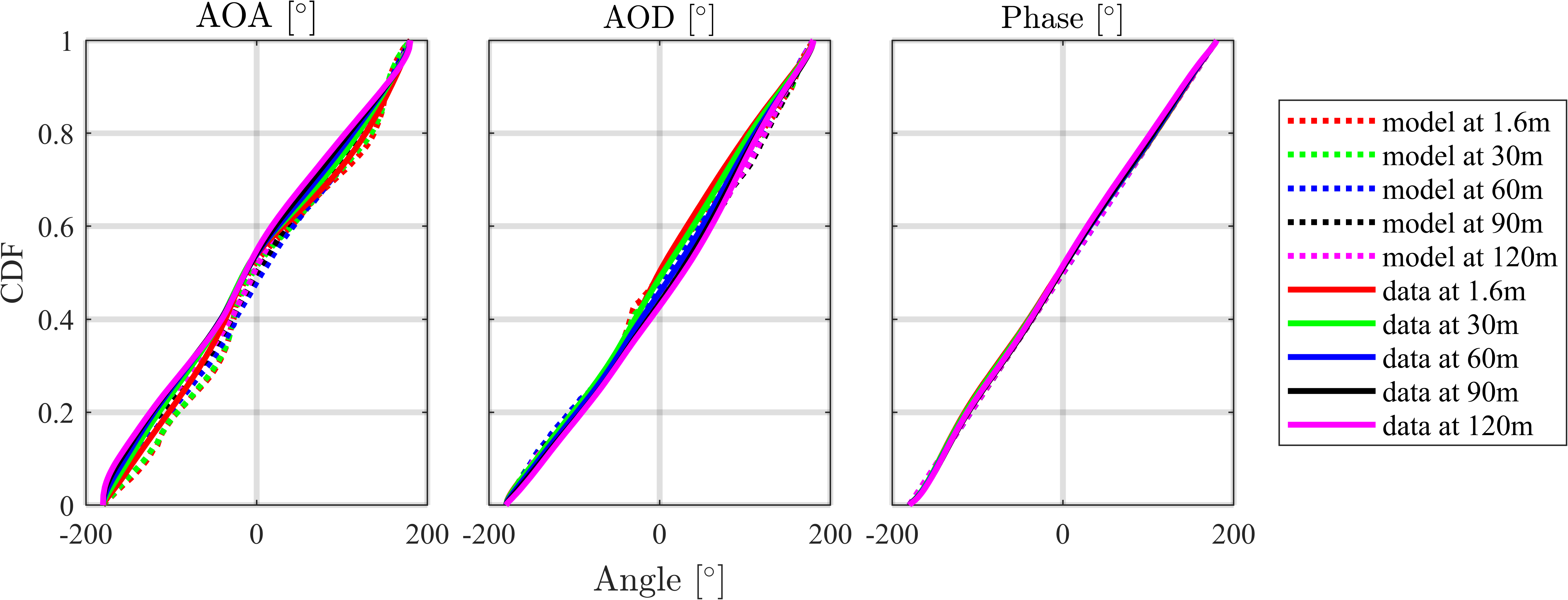} 
\caption{CDFs of AOA, AOD and arrival phase}
\label{fig:azm_phase}
\end{figure}

As shown in Fig~\ref{fig:azm_phase},  the distributions of AOA, AOD, and arrival phase are \emph{uniform} at each height due to the conditionality of the 2D distance.
\begin{figure*}[!t]
\centering
\subfloat[][]
{\includegraphics[width =0.41\columnwidth]{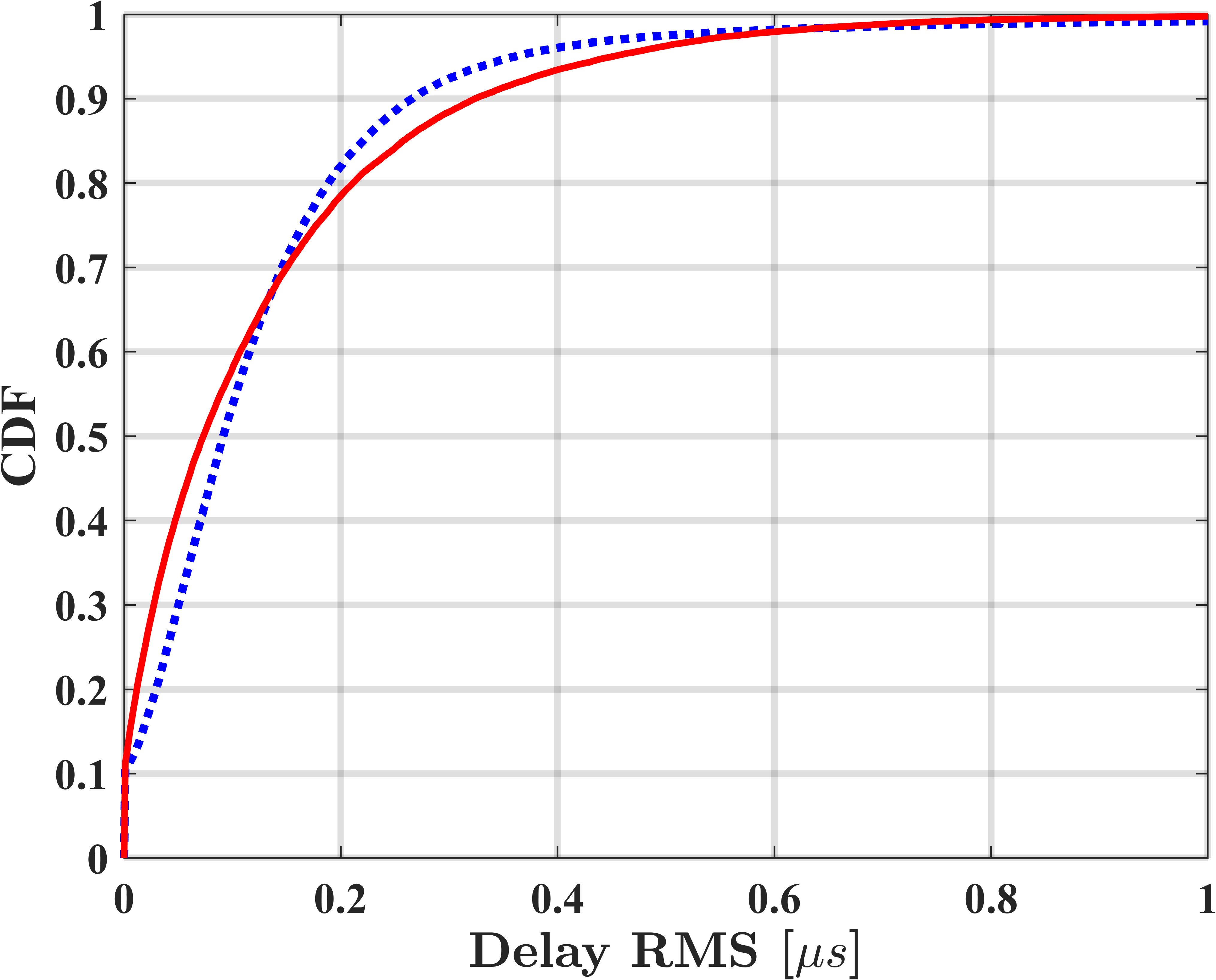} \label{fig:rms_delay}}
\subfloat[][]
{\includegraphics[width =0.38\columnwidth]{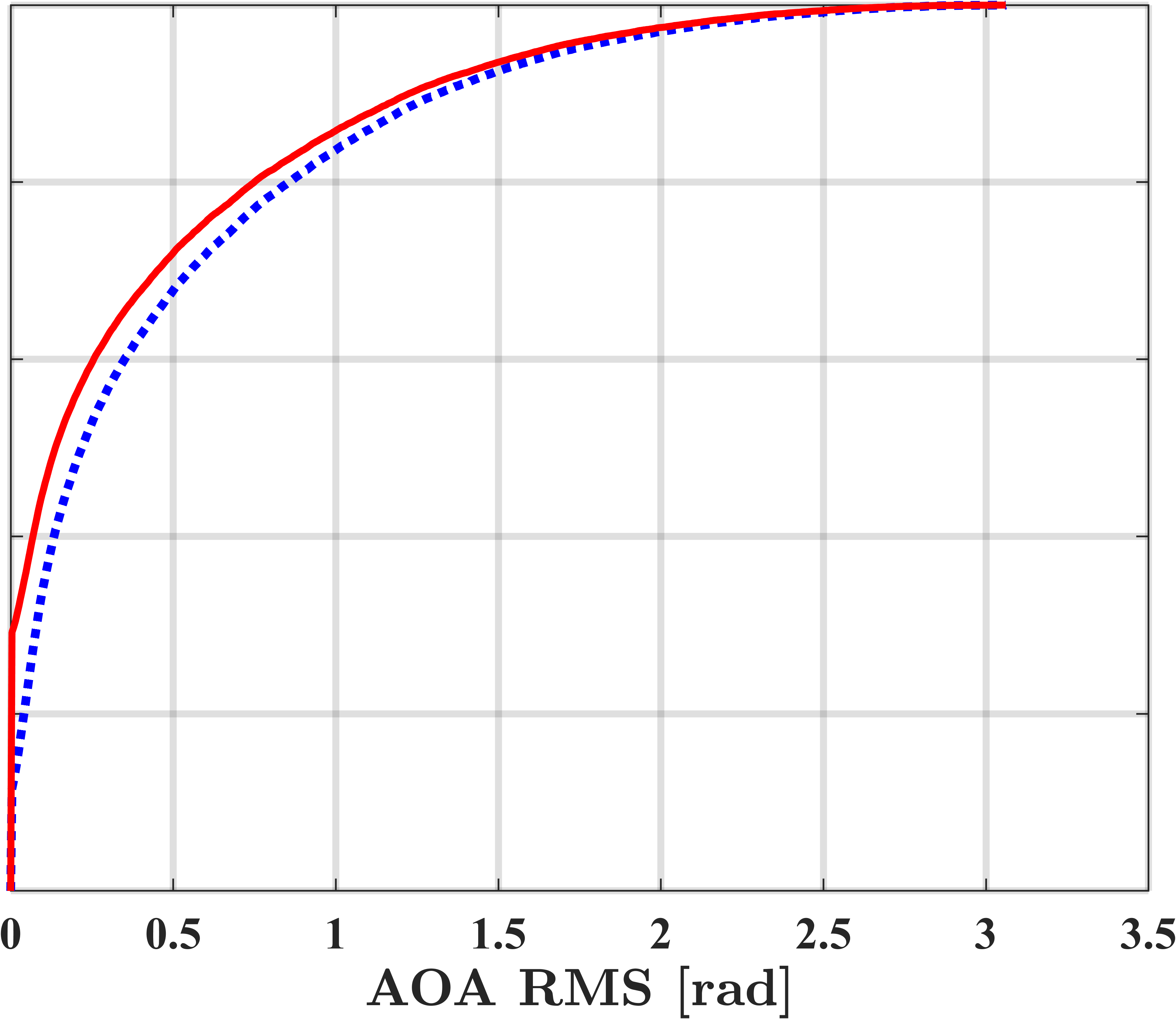} \label{fig:rms_aoa}}
\subfloat[][]
{\includegraphics[width =0.38\columnwidth]{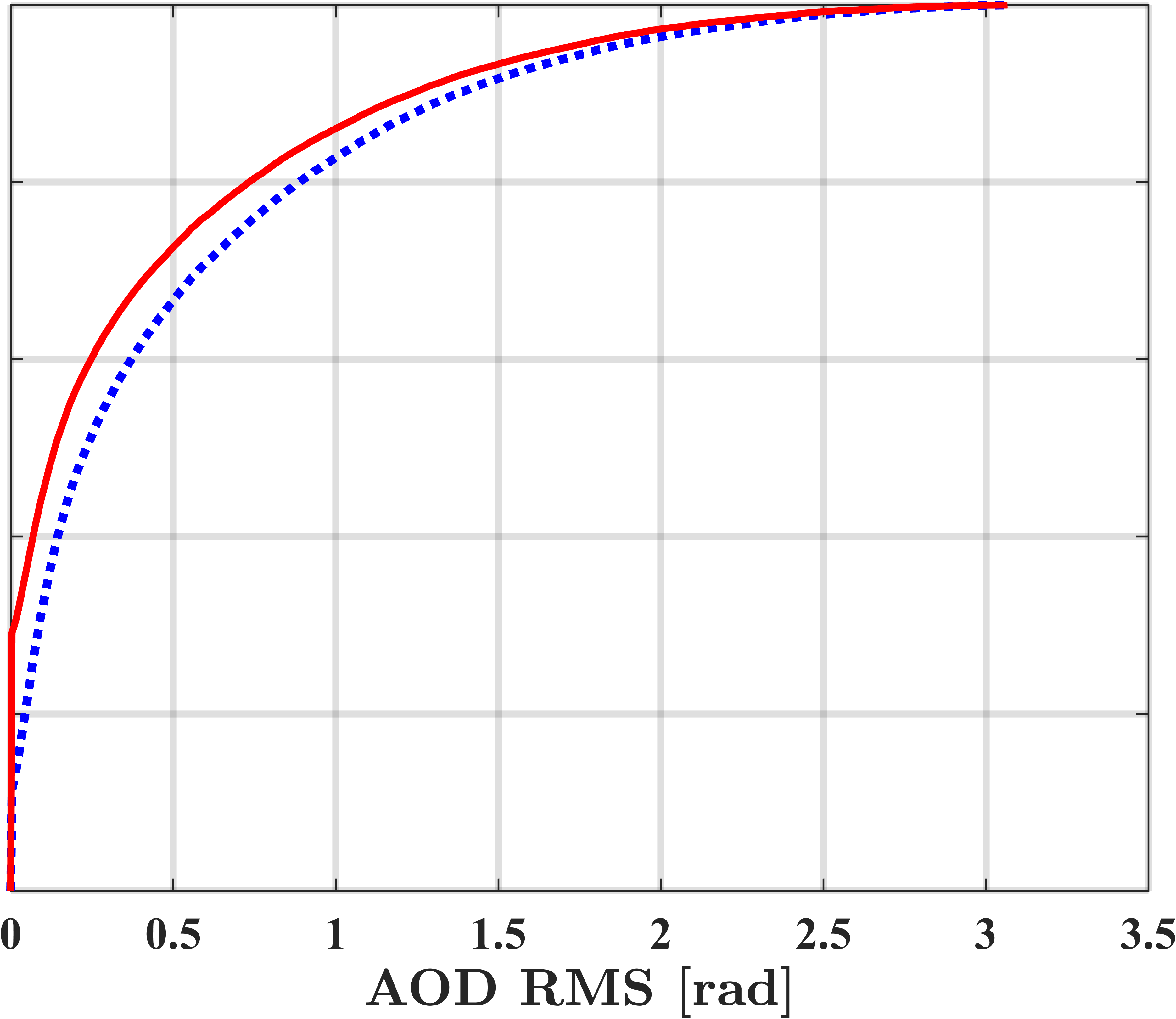} \label{fig:rms_aod}}
\subfloat[][]
{\includegraphics[width =0.38\columnwidth]{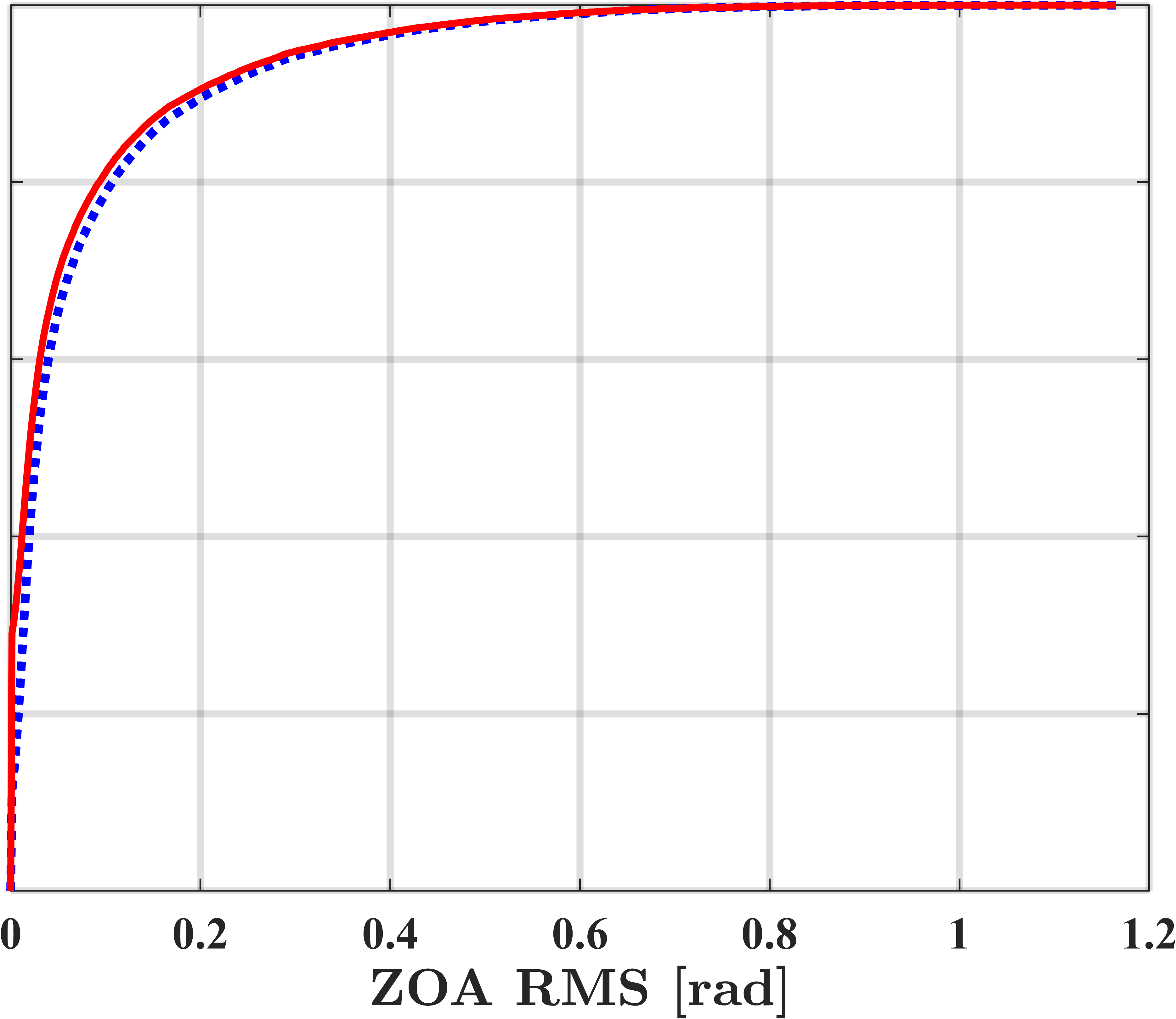} \label{fig:rms_zoa}}
\subfloat[][]
{\includegraphics[width =0.38\columnwidth]{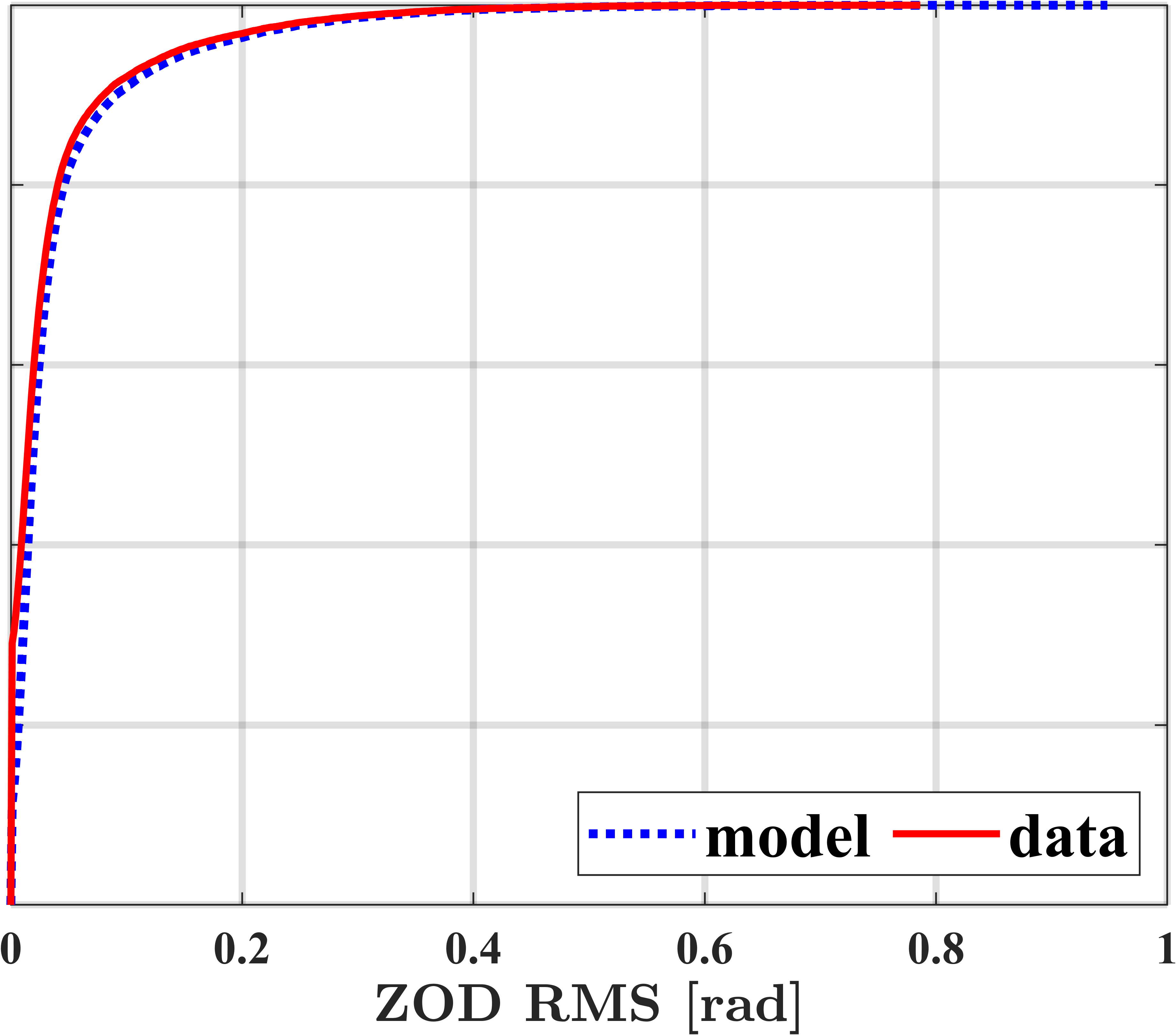} \label{fig:rms_zod}}
\caption{RMS CDFs for (a) delay, (b) AOA, (c) AOD, (d) ZOA, and (e) ZOD}
\label{fig:rms_cdf}
\end{figure*}

As the last evaluation of the model, we briefly discuss the root mean square (RMS) spread for the delay, azimuth, and zenith angles, which represents the degree of sparsity or richness of the multipath channels. According to \cite{ITURMS2017}, the general RMS spread values for a multipath feature $\boldsymbol{d}$ are calculated as follows:

\begin{center}
\begin{align*}
    d_{r m s} =\sqrt{\frac{\sum_k\left(d_k-d_m\right)^2 P_k}{\sum_k P_k}},~d_m =\sum_k \frac{d_k P_k}{\sum_k P_k}
\end{align*}
\end{center}
where $P_k$ is path gain, $d_k$ is channel parameters such as \emph{excess delay}, arrival and departure angles at k'th path.
 RMS values are instrumental metrics in dense urban environments where there are many local scatters. We consider the RMS values of delay, AOA, AOD, ZOA, and ZOD when the RX height is $\SI{1.6}{m}$, as at lower altitudes rich scatterings can occur.  
The statistical distributions of the RMS values are shown in Fig.~\ref{fig:rms_cdf} as CDF for AOA, AOD, ZOA, ZOD and delay. As shown, we observe that the distributions are very closely aligned with those of the original data. 
 These results imply that the WGAN-GP statistically well captures the multipath characteristics that reflect rich or sparse scatterings.

\section{Conclusion}
In this work, we introduce a simplified method for building GBSM given specific conditional constraints based on channel parameter data obtained from ray-tracing simulation. Because building a theoretical GBSM that captures all the statistics of local scatters is very challenging, generative neural networks are considered. Specifically, WGAN-GP is adopted in this paper. We propose methods to convert all channel parameters to images to reduce the complexity of WGAN-GP training and implementation. After training the model, we confirm that the statistics of all the channel parameters of WGAN-GP are well matched with the distributions of the original data.

Thus, we argue that given wireless channel data from any geometry or a specific site, generative models can be leveraged with the proposed channel image methods for GBSM.  Depending on the accuracy constraints of modeling and implementation complexities,  other generative models, such as VAE and the diffusion model, can be employed. Finally, using the proposed channel images, other wireless conditions, such as multi-carrier frequencies and different scenarios of environments, will be readily implemented to assess and compare the network performances.

\bibliographystyle{IEEEtran}
\bibliography{bibl}

\vfill

\end{document}